\documentclass[twocolumn]{aastex631}

\usepackage{url}
\usepackage{epsfig}
\usepackage{graphicx}
\usepackage{xcolor}
\usepackage{hyperref}
\usepackage{threeparttable}
\usepackage{multibib}
\usepackage{amssymb,amsmath}

\begin{document}

%\title{Attempting to Link FRBs and Pulsars Through High-Energy Distributions}
%\title{A tentative link for FRBs and pulsars with wide energy distributions}
%\title{Exploring a tentative link between FRBs and pulsars with broad energy distributions}
%\title{Exploring a tentative link between FRBs and pulsars with broad energy distributions and applications for nearby FRB survey strategies}
%\title{\textcolor{blue}{Exploring FRB$-$pulsar similarities and implications for nearby FRB surveys}}
%\title{\textcolor{blue}{Investigating the flat tails in burst energy distribution and implications for nearby FRB surveys}}
\title{Flat tails in FRB and pulsar energy distributions: implications for optimizing nearby FRB surveys}

\author{S.B. Zhang}
\affiliation{Purple Mountain Observatory, Chinese Academy of Sciences, Nanjing 210023, China}
\affiliation{CSIRO Space and Astronomy, Australia Telescope National Facility, PO Box 76, Epping, NSW 1710, Australia}

\author{G. Hobbs}
\affiliation{CSIRO Space and Astronomy, Australia Telescope National Facility, PO Box 76, Epping, NSW 1710, Australia}

\author{S. Johnston}
\affiliation{CSIRO Space and Astronomy, Australia Telescope National Facility, PO Box 76, Epping, NSW 1710, Australia}

\author{S. Dai}
\affiliation{CSIRO Space and Astronomy, Australia Telescope National Facility, PO Box 76, Epping, NSW 1710, Australia}

\author{Y. Li}
\affiliation{Purple Mountain Observatory, Chinese Academy of Sciences, Nanjing 210023, China}

\author{J. S. Wang}
\affiliation{Max-Planck-Institut f\"ur Kernphysik, Saupfercheckweg 1, D-69117 Heidelberg, Germany}

\author{X. Yang}
\affiliation{Purple Mountain Observatory, Chinese Academy of Sciences, Nanjing 210023, China}
\affiliation{School of Astronomy and Space Sciences, University of Science and Technology of China, Hefei 230026, China}

\author{X. F. Wu}
\affiliation{Purple Mountain Observatory, Chinese Academy of Sciences, Nanjing 210023, China}
\affiliation{School of Astronomy and Space Sciences, University of Science and Technology of China, Hefei 230026, China}

\author{L. Staveley-Smith}
\affiliation{International Centre for Radio Astronomy Research, University of Western Australia, Crawley, WA 6009, Australia}
\affiliation{ARC Centre of Excellence for All Sky Astrophysics in 3 Dimensions (ASTRO 3D)}



\begin{abstract} %up to 250 words
Fast radio bursts (FRBs) are energetic, short-duration radio pulses of unclear origin. 
To explore effective survey strategies for detecting FRBs from nearby globular clusters (GCs), we investigate the burst energy distribution, which has a strong influence on the detection rate.
%To explore effective survey strategies for detecting FRBs from nearby globular clusters (GCs), which strongly influenced by the burst energy distribution, 
We re-analyze FRBs and pulsars exhibiting broad energy distributions by fitting their high-energy tails with power-law models.
Two cosmological repeating FRBs (FRB 20201124A and FRB 20220912A), one nearby FRB (FRB 20200120E), and two pulsars (RRATs J1846$-$0257 and J1854+0306), exhibit power-law indices of $\alpha \gtrsim -1$, suggesting that their bright pulses contribute significantly to the total radio pulse energy. The brightest bursts from these sources can be fitted with a power-law model ($\alpha_{\rm Bri} = -0.26 \pm 0.05$), suggesting that an extremely flat index is required to naturally derive high-luminosity FRBs from low-luminosity sources.
We present detailed survey strategies for FAST, MeerKAT and Parkes cryoPAF in the search for FRBs in nearby GCs using different power-law indices, recommending targets for observation. We suggest that combining observations with FAST ($\sim3$ hours) and Parkes cryoPAF (10$-$20 hours) is practicable for discovering new FRBs in nearby GCs.
\end{abstract}


%% We recommend that authors also use the natbib ~\citepp
%% and ~\citept commands to identify citations.  The citations are
\clearpage

\section{Introduction} \label{sec:intro}

%FRB intro
Fast radio bursts (FRBs) are immensely energetic radio pulses with durations ranging from microseconds to seconds~\citep{Lorimer07, Snelders23, Chime22_subs}. Among the nearly one thousand known FRB sources, fewer than 10\% exhibit repeat bursts~\citep{CHIME21, Spitler16, CHIME23}. 
%
%FRB origin
The origin of FRBs remains unclear, but the only two nearby events offer clues to their engines: the apparently one-off burst FRB 20200428 from the Galactic magnetar SGR~1935+2154~\citep{Bochenek20, CHIME20} and the repeating FRB 20200120E, localised to a globular cluster (GC) in the M81 galactic system at a distance of 3.63\,Mpc~\citep{Bhardwaj21,Kirsten22}. 
While these two events suggest connections to distinct environments, young magnetars and ancient star populations, respectively, both have propelled FRB theories involving compact objects, especially neutron stars, as potential sources~\citep{Zhang23_FRBModel}. 
%{\bf talking some specified model?}

%require for nearby FRBs
Nearby FRBs are crucial in understanding FRBs, as they enable precise localisation to identify their surroundings and deep multiwavelength follow-up to resolve their mechanisms.  
Although CHIME, the current most efficient FRB hunter, has been operating since 2018 and can discover FRBs daily~\citep{CHIME21}, only one repeating FRB with a distance $\leq$ 100\,Mpc has been detected~\citep{Bhardwaj21}, indicating that a blind survey is challenging for providing a sufficient sample of nearby repeating FRBs. 
As the only nearby repeating FRB~20200120E has been localised to a GC~\citep{Kirsten22}, \citet{Kremer23} presented a dedicated FRB survey of GCs in nearby galaxies, with M87 being identified as the most promising target, where high-sensitivity radio telescopes like 
%FAST~\citep{Jiang20} or MeerKAT~\citep{Bailes20} 
FAST or MeerKAT
have a 90\% probability of detecting a globular cluster FRB within approximately 10 hours of observation.

%The key facts of nearby FRB search from GCs 
However, it is notable that the detection rate of nearby FRBs from such dedicated surveys is significantly influenced by the burst energy distribution~\citep{Kremer23}. The nearby radio bursts expected to be detected by high-sensitivity telescopes such as FAST~\citep{Jiang20} would have relatively low energies, even comparable to pulsar single pulses.
The expected event rate from surveys such as those proposed by~\citet{Kirsten22} assumes a consistent distribution from bright to faint target sources. Therefore, it is important to investigate the relationship between bright sources, such as cosmological FRBs, and relatively faint sources, such as nearby FRBs or even bright pulsar single pulses.
%The expected event rate includes the detection of both bright FRB bursts and relatively faint bursts.
%Therefore, investigating the FRB energy distributions and their relationship with relatively low-energy radio sources is important to optimize nearby FRB surveys.
%From FRB to RRATs

%Broad energy distributions of FRBs 
%Talk about FRB 20200120E first.
FRB 20200120E stands as the sole nearby repeating FRB at a distance of approximately 3.63\,Mpc that enables detailed energy distribution analysis.
%burst strom
Observations with the Effelsberg telescope recorded 65 bursts~\citep{Nimmo23,Kirsten22} with fluences ranging from 0.04–0.71\,Jy\,ms at 1.4\,GHz, including a remarkable ``burst storm'' event of 53 bursts detected within 40 minutes. The burst storm exhibited a steep energy distribution characterized by a power-law tail with index of $\sim -2.39$~\citep{Nimmo23}. 
%bright burst
However, a recent observation have detected an exceptionally bright burst (fluence $\approx 30$\,ms at 1.1-1.7\,GHz), 42 times brighter than previous detection at 1.4 GHz, suggesting a flatter high-energy tail with index of approximately $-0.98$~\citep{Zhang24_M81}.

%cosmological FRBs.
%Add papers of FRB121102, which also indicate an index flatter than -1 (should double check the Li2021 data, which seems a bit different from our J1913+1330 paper fitting results),
%(For 121102 in particular Cruces+2021, Hewitt+2022, Jahns+2022, Li+2021, all find power-law indices around -1.)
The typical energy of cosmological repeating FRBs~\citep{Li21, Xu22, Chime20_Li} is approximately four orders of magnitude greater than that of FRB~20200120E~\citep{Nimmo23}.
The study of energy distributions in cosmological FRBs began with FRB~20121102A, the first known repeater~\citep{Spitler16}. 
As summarized by~\citet{Wang19_FRBDis}, the early repeating samples of FRB~20121102A showed
%presenting
cumulative energy distributions of power-law tails with indices from $~-0.6$ to $-0.8$\footnote{It is noteworthy that their definition of the indices differs from that in our study, we have converted their values to match our definition.}. Later analysis with larger samples also indicate flatter high energy tails with indices larger than $-1.0$~\citep{Cruces21, Hewitt22, Jahns23, Li21}. 
Although monitoring of other active repeaters such as FRB~20180916B and FRB~20190520B showed steep indices~\citep[($\alpha < -2.0$)][]{Chime20_Li,Zhang22RAA}, recent studies of two hyperactive repeaters, FRB 20201124A and FRB 20220912A, both detected extreme bright pulses by ultra-long monitoring ($> 2000$ hours), revealing breaks in their burst energy distributions, with a flattening of the power-law slope ($\gtrsim -1$) at higher energies~\citep{Kirsten23,Ould-Boukattine24}.
These flatter indices at high energies mirror FRB 20200120E’s bright burst properties.   

%could refer the latest Yuannan paper (KJ). also including FRB121120, FRB240114 in the paper discussion. 
%add more detailed discussion for all active FRBs, including "active sources to consider might be the FRBs 121102, 180916B, 190520B, 220529, 220912 and possibly the more recent 240114"
Furthermore, the latest rate analysis of the new detected hyperactive repeater FRB~20240114A also indicates a potential flattening at higher energies~\citep{Huang25}.
%

%pulsars
The typical energy of pulsars~\citep{Manchester05} is about twelve orders of magnitude smaller than that of cosmological repeating FRBs~\citep{Li21, Xu22, Chime20_Li}.
The broad energy distributions observed in FRBs are also atypically for single pulses from pulsars~\citep{Burke-Spolaor12}. Intensity variations exceeding an order of magnitude are usually attributed to the rare phenomenon of ``giant pulses''~\citep{Kuzmin07}, reported in only about ten~\citep{Wang19} of the up to 4000 known pulsars\footnote{ATNF Pulsar Catalogue v2.6.1: \url{https://www.atnf.csiro.au/research/pulsar/psrcat}}~\citep{Manchester05}.
%Giant pulses
Pulsar giant pulses normally exhibit steep energy distributions. For instance, J0534+2200 (the Crab pulsar) and J0540$-$6919 are known for producing giant pulses with the largest excesses relative to their normal emissions~\citep{Bera19, Geyer21, Wang19}. 
The Crab pulsar (J0534+2200) demonstrates a single power-law distribution with $ \alpha \sim -3.0$ across three decades of fluence. 
Similarly, J0540-6919 shows a cumulative distribution with $ \alpha = -2.75 \pm 0.02 $.
%%RRATs
However, recent observations of another kind of special pulsars~\citep{Zhang24_1913, Zhang24_highB}, rotating radio transients (RRATs), characterized by sporadic radio pulses, have also shown broad energy distributions. Although one of which (J1913+1330) present steep tails~\citep{Zhang24_1913}, two other RRATs (J1846$-$0257 and J1854+0306) have also exhibited power-law-like tails at high energies with similarly flat indices~\citep[$\gtrsim -1$; ][]{Zhang24_highB}.
These indices match the flatter high-energy regimes of FRB~20200120E and cosmological repeaters like FRB 20201124A and FRB 20220912A.

The detection of a bright burst from FRB 20200120E~\citep{Zhang24_M81} bridges the gap between nearby and cosmological FRBs, and the detection of radio bursts with varying luminosities from SGR~1935+2154~\citep{Bochenek20, CHIME20, Kirsten21, Good2020ATel, Zhu23} connects the energy range of nearby FRBs to that of pulsars. No clear gap exists between the estimated isotropic-equivalent energy releases of various radio pulses~\citep{Zhang24_M81}, making it natural to explore the burst energy distribution relationship across radio sources with various energy releases.

Recent studies have shown that flat tails in pulse energy distributions are observed across cosmological FRBs, the nearby FRB 20200120E, and certain pulsars. However, analysis of different sources has employed various methods: some studies derived power-law indices from cumulative distributions, while others used differential distributions. Energy calculations have varied depending on whether they incorporated central frequency, bandwidth, or burst frequency extent scaling. Additionally, the choice of distribution bins and fitting tools can affect the obtained index values. 
In this study, we present a consistent analysis of the high-energy tails in the energy distributions of FRBs and pulsars using a unified methodological approach.
Section~\ref{sec:data} describes the source selection and data processing, while Section~\ref{sec:res} and~\ref{sec:dis} presents and discusses the results. We conclude in Section~\ref{sec:con}.

%\clearpage

\section{Source selection and data processing} \label{sec:data}

\subsection{Source selection}

%%data sections
%why these sources, the low-energy burst exists 
%This letter focuses on the high-energy distribution tails of various radio-pulsing sources, emphasizing those that exhibit both low- and high-energy emissions.
%sources with flat tails
Our selection prioritizes sources that exhibit flat high-energy tails in their burst energy distributions. Among FRBs, we include four cosmological sources with demonstrated power-law indices of $\alpha \gtrsim -1$ at high energies: FRB~20121102A, FRB~20201124A, FRB~20220912A, and FRB~20240114A. We also include the nearby FRB 20200120E, which shows a tentative high-energy tail ($\alpha \sim -0.98$), a key target of our study.
For pulsars, we focus on J1846$-$0257 and J1854+0306, two RRATs exhibiting flat tails, in contrast to the steep energy distributions typically observed in giant pulse pulsars. For comparison, we selected the well-studied giant pulse pulsars J0534+2200 (the Crab pulsar) and J0540$-$6919, known for producing giant pulses with the largest excesses relative to their normal emissions.
However, detailed burst properties for FRB~20220912A and FRB~20240114A are currently unavailable, limiting our ability to conduct comprehensive energy distribution analyses. Similarly, FRB~20200120E has only one documented ultra-bright burst, which restricts statistical analysis. Therefore, our quantitative energy distribution analysis focuses on the data of FRB~20121102A~\citep[][; the largest available sample]{Li21} and FRB~20201124A~\citep[][; the sample containing the brightest burst]{Kirsten23} among the FRB sources, J1846$-$0257 and J1854+0306~\citep{Zhang24_highB} among the pulsar sources, and J0534+2200~\citep[][the sample containing the largest number of bright bursts]{Bera19} and J0540$-$6919~\citep[][; the most extensive sample]{Geyer21} as giant pulse comparison sources.

\subsection{Data processing}

%%data processing
%the fluence data, equivalent to the isotropic-equivalent special energy   
To compare the energy distributions of different radio pulses, we calculated the specific isotropic-equivalent energy using~\citep{Zhang18}: 
\begin{equation}
\centering
E_S=\frac{4\pi D_{L}^2}{1+z} \,F,
\label{equ:energy}
\end{equation}
where $D_{L}$ is the luminosity distance, $z$ is the redshift, and $F$ is the reported pulse fluence from the burst samples described above. 
The energy values for each source were then normalized to their brightest bursts, as our analysis focuses on comparing the high-energy tails.
%bins, density

Since only the cumulative energy distribution tail is available for the only nearby repeating FRB 20200120E, we used the power-law indices ($\alpha$) derived from the cumulative energy distribution in this study, which is given by:  
\begin{equation}
\centering
{\rm log}_{10} N (\ge E) = C + \alpha \, {\rm log}_{10} E,
%N (\ge E) = C + \alpha E,
\label{equ:cum_ind}
\end{equation}
where $N (\ge E)$ represents the predicted number of bursts with specific energy above $E$. 
%
%%%wahy used the cumulative? explain this
%Since only one bright burst has been detected from FRB 20200120E~\citep{Zhang24_M81} and detailed burst properties for FRB 20220912A are currently unavailable~\citep{Ould-Boukattine24}, we were unable to reprocess their high-energy distributions. 
%To compare our results with their reported distributions, which are cumulative, 
%We performed both cumulative (hereafter denoted by index $\alpha$) and differential (hereafter denoted by index $\beta$) distribution analyses, 
In our distribution analysis, we adjusted
the bins for each source to cover a similar range in log-10 space.
%
%We also provide a differential distribution analysis.
%although the sample size for some studied sources are limited, and the fitted uncertainties are relatively large.   
%
To enable comparisons between different sources, we defined a density that is calculated as the number of bursts per bin divided by the total number of detected bursts. 

%%%threshold for fit? 90%? for 1913, second peak?

%fitting: curve_fit
%add uncertainty
For data of pulses from a single source (i.e., Figure~\ref{fig:indi}), we considered only simple symmetric uncertainties in the counts, and fitted the power-law-like tails using
the Python package~\textsc{curve\_fit} function from the~\textsc{SciPy} library~\citep{SciPy}.
The cutoff point for fitting was determined by identifying the location where the standard error was minimized.
%complicated uncertainty
For data of pulses from different sources (i.e., Figure~\ref{fig:rate}), we performed the fitting within a Bayesian framework using Markov Chain Monte Carlo (MCMC) sampling, via the~\textsc{emcee} Python package~\citep{emcee2013}. The fitting was conducted in log-log space using an asymmetric Gaussian likelihood.
%For data with asymmetric uncertainties, we performed the fitting within a Bayesian framework using Markov Chain Monte Carlo (MCMC) sampling, via the~\textsc{emcee} Python package~\citep{emcee2013}, similar to the approach used by~\citet{James19}.

%\begin{equation}
%\log \mathcal{L}(\theta) = -\frac{1}{2} \sum_{i=1}^{N} \left[ \frac{(y_i - m_i(\theta))^2}{\sigma_{i}^{2}} + \ln(2\pi \sigma_{i}^{2}) \right]\\
%\end{equation}
%where
%\begin{equation}
%\sigma_{i} = 
%\begin{cases}
%\sigma_{i,\text{upper}} & \text{if } y_i > m_i(\theta) \\
%\sigma_{i,\text{lower}} & \text{if } y_i \leq m_i(\theta)]
%\end{cases}
%\end{equation}

\section{Results}  \label{sec:res}

%\subsection{Cumulative distributions at high energy of various radio sources}

%%%%add a figure for individual sources fitting, including error-bar
\begin{figure*}[!htb]
\centering
\begin{tabular}{lll}
\includegraphics[width=0.28\linewidth]{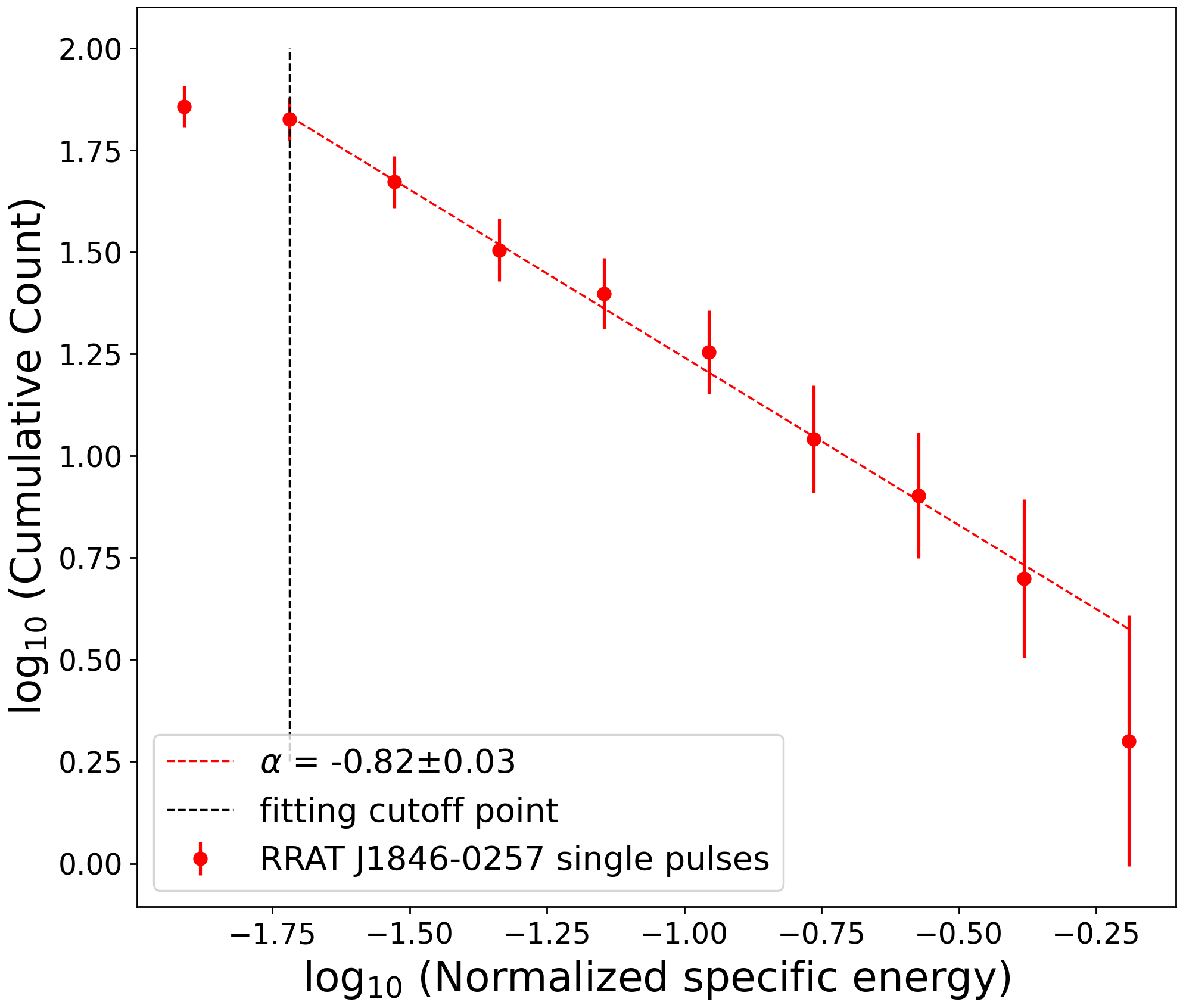} &
\includegraphics[width=0.28\linewidth]{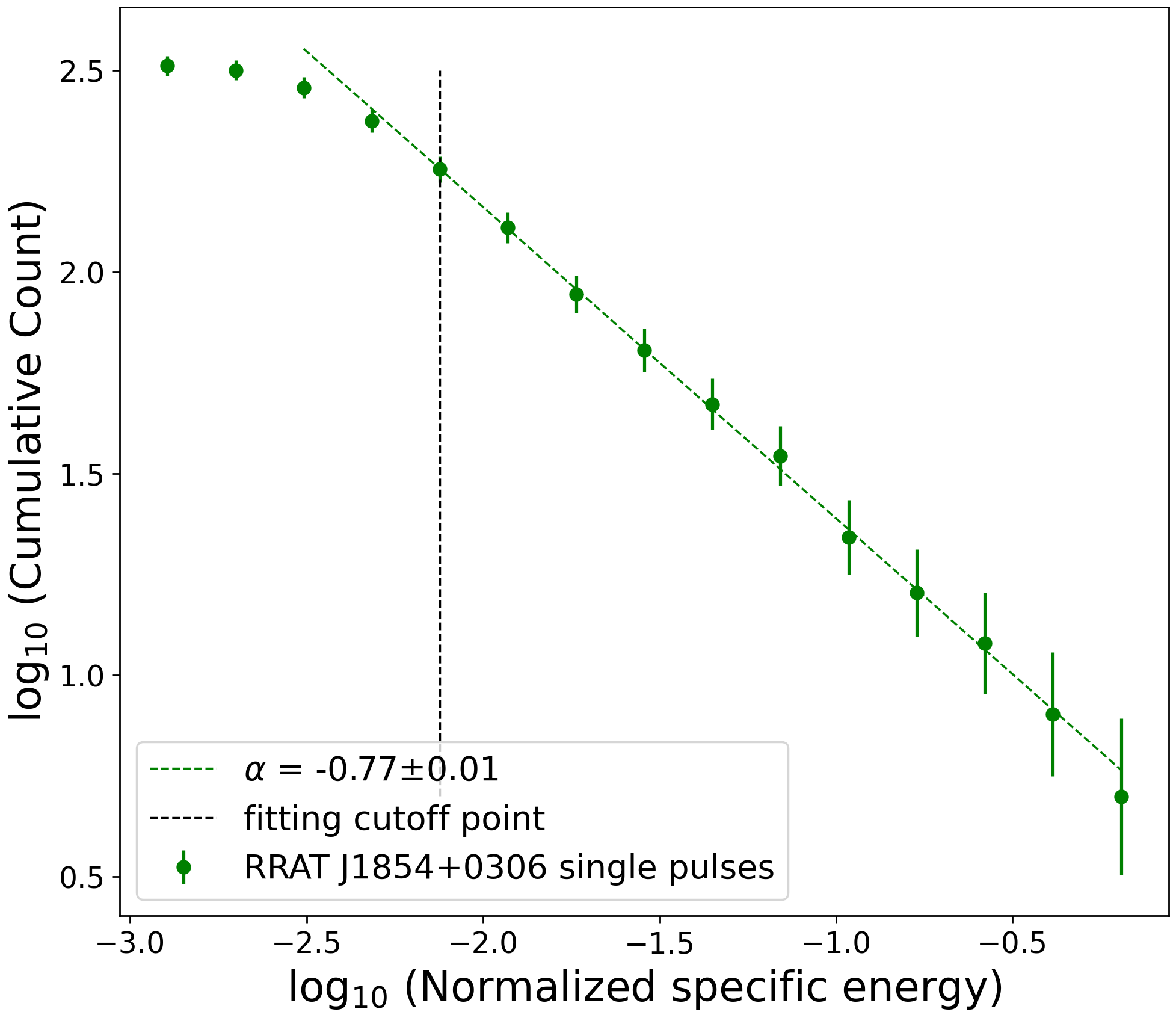} &  
\includegraphics[width=0.28\linewidth]{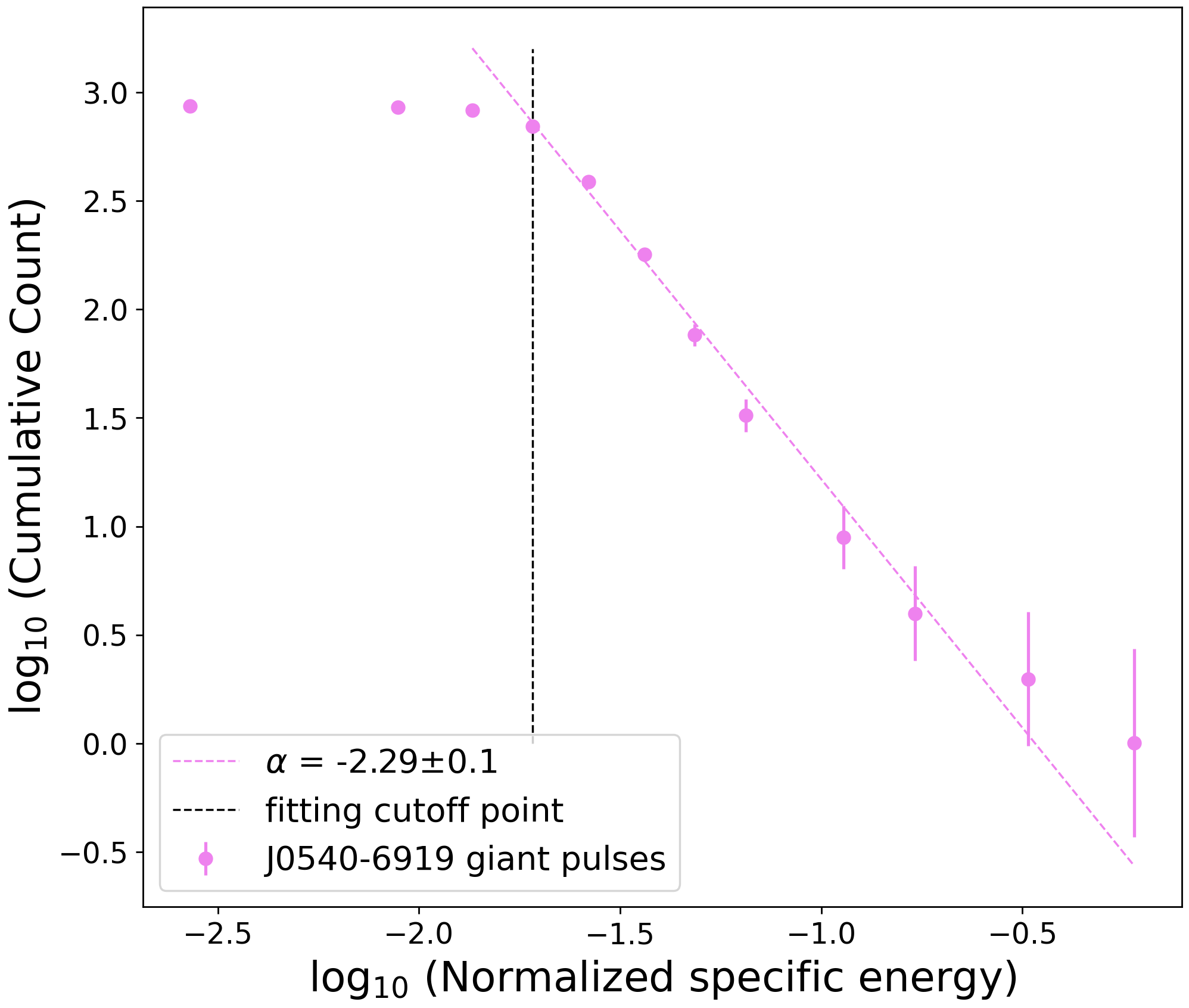} \\
\includegraphics[width=0.28\linewidth]{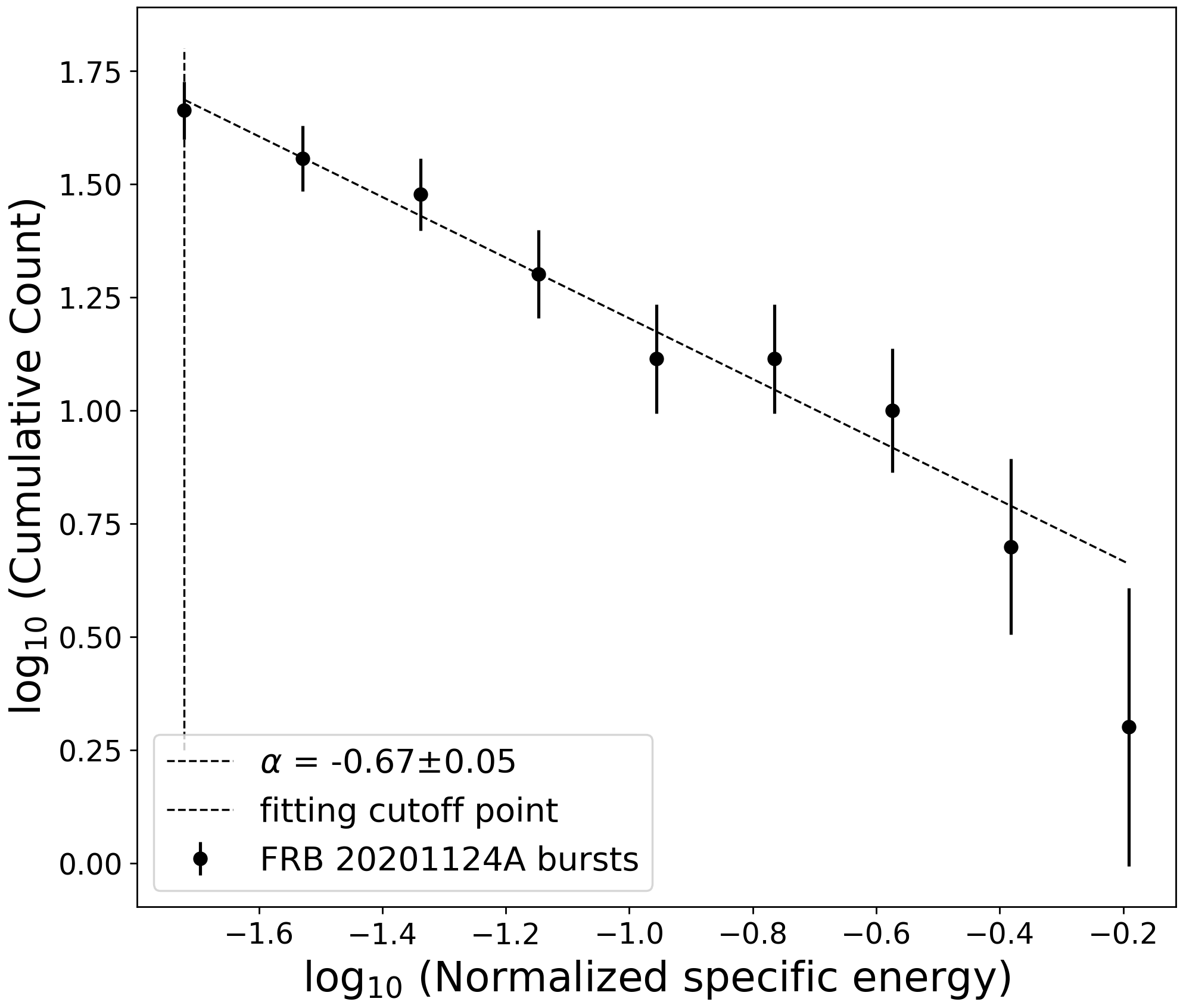}  &
\includegraphics[width=0.28\linewidth]{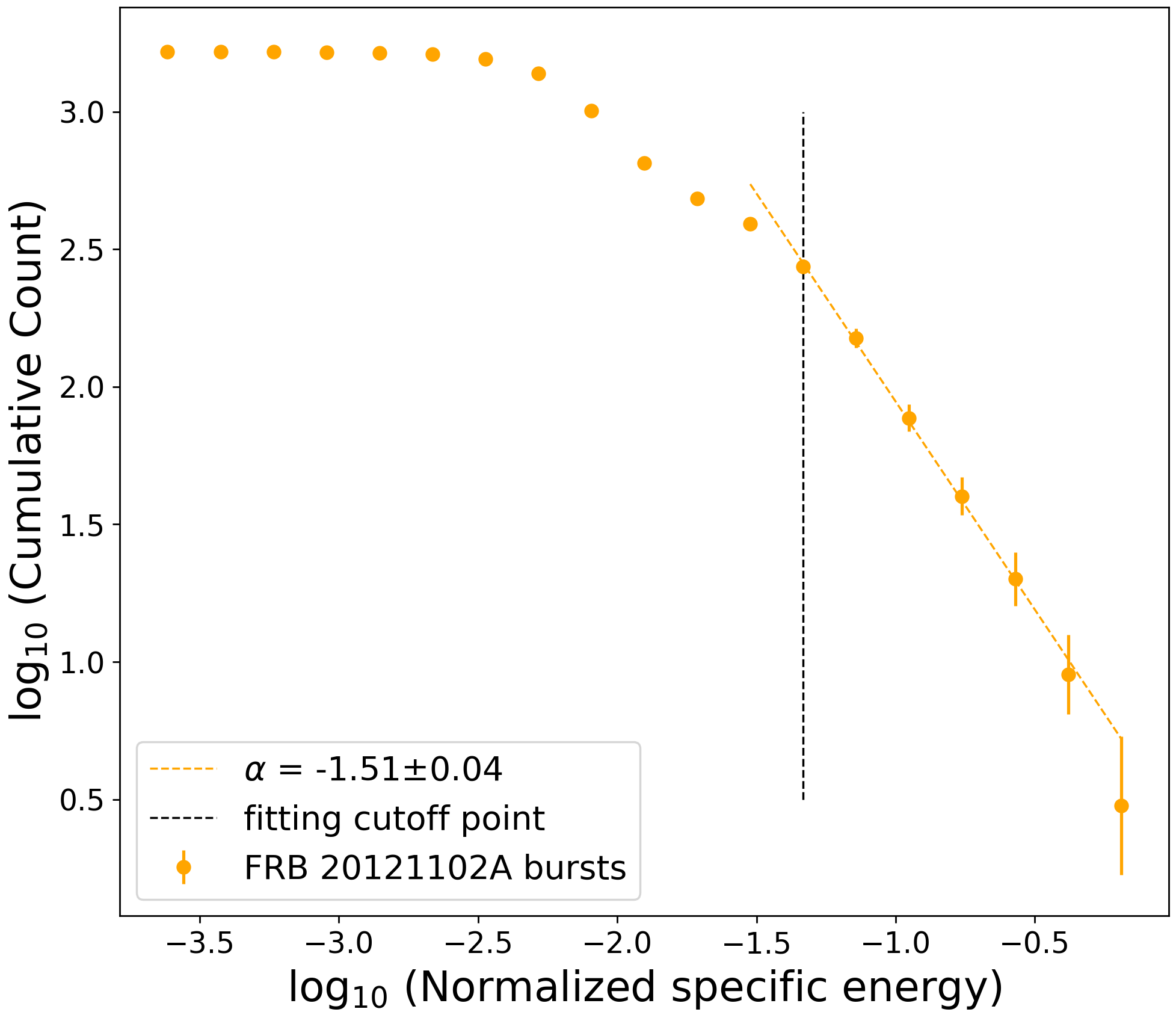}  &
\includegraphics[width=0.28\linewidth]{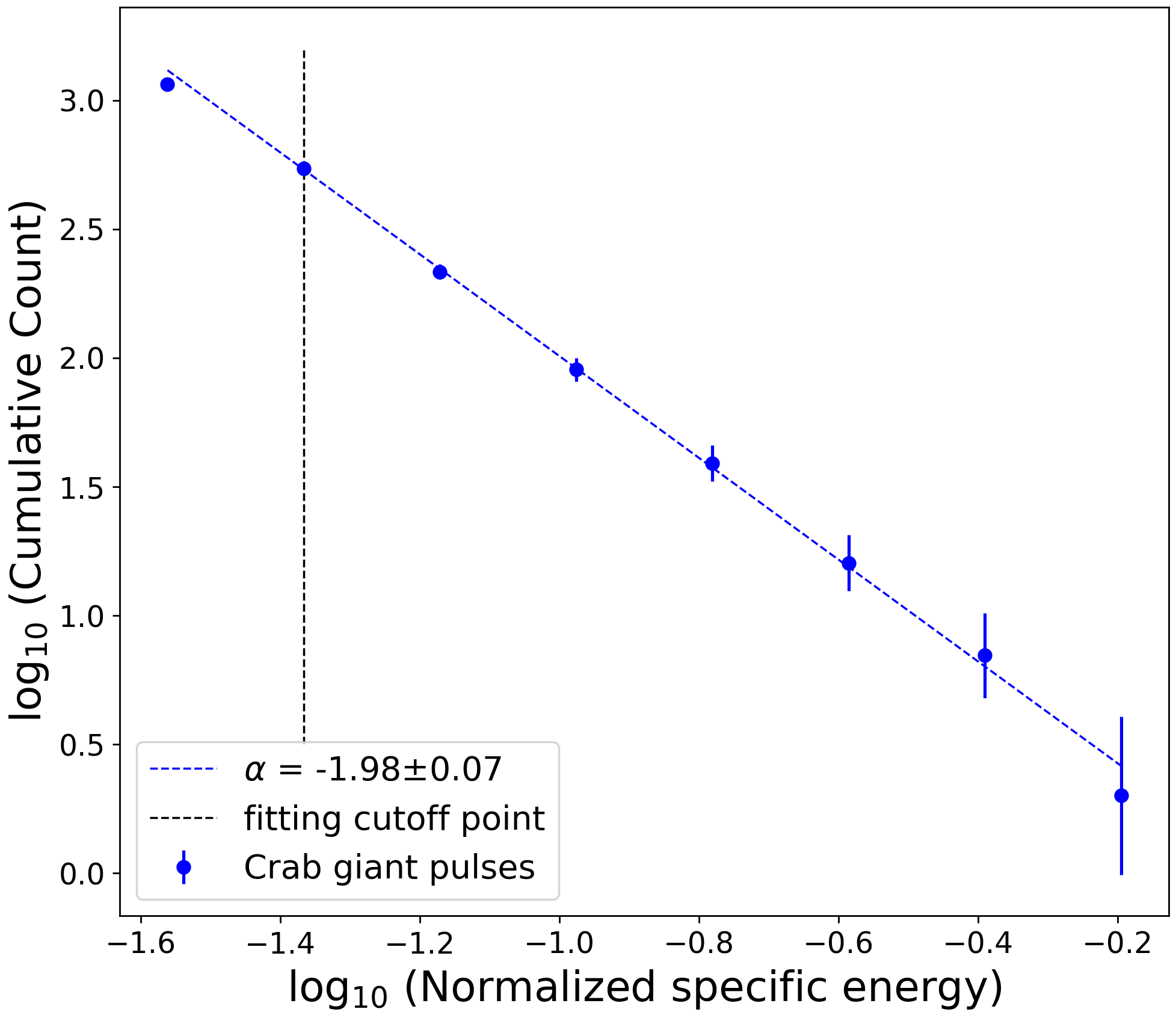} \\

\end{tabular}
\caption{Cumulative specific energy distributions of high-energy bursts from FRB 20201124A and FRB~20121102A, single pulses from RRATs J1846$-$0257 and J1854+0306, and giant pulses from the Crab pulsar (J0534+2200) and J0540$-$6919. The energy values for each source are normalized to their respective brightest bursts. Each high-energy tail is fitted with a simple power law model. 
The fitting cutoff points are indicated by black dashed lines.}
\label{fig:indi}
\end{figure*}

Figure~\ref{fig:indi} presents the cumulative specific energy distributions of high-energy bursts from the selected sources with relatively sufficient burst counts, most of which can be well-fitted by a simple power-law model.
%
%The high-energy tail of the distribution for RRAT J1913+1330 is forcefully fitted using a power-law function, although the fit is not as good as for the other sources, in order to enable comparison. 
%indices
Specifically, the fitted power-law indices ($\alpha$) for FRB 20201124A (46 bursts\footnote{For events detected simultaneously by multiple telescopes, we selected bursts from the observation with the largest bandwidth.}), RRATs J1846$-$0257 (72 bright pulses) and J1854+0306 (325 bright pulses) are $-0.67 \pm 0.05$, $-0.82 \pm 0.03$ and $-0.77 \pm 0.01$, respectively, all of which are $\gtrsim -1$.
%
%In contrast, the indices for the Crab pulsar ($-1.98 \pm 0.07$, 1955 giant pulses), J0540$-$6919 ($-2.29 \pm 0.07$, 865 bright single pulses) and RRAT J1913+1330 ($-3.04 \pm 0.29$, 1955 individual pulses) are significantly steeper, at $\lesssim -2$.
In contrast, the indices for the Crab pulsar ($-1.98 \pm 0.07$, 1955 giant pulses) and J0540$-$6919 ($-2.29 \pm 0.1$, 865 bright single pulses) are significantly steeper, at $\lesssim -2$.
%FRB 2012102A
The index for FRB~20121102 is $-1.51 \pm 0.04$ (1652 bursts), falling between the above two categories.
%add results of FRB20200120E and FRB 20220912A.
Although the available data for bursts from FRB 20200120E and FRB 20220912A are insufficient to reprocess their high-energy distributions, the reported fitted power-law indices from cumulative distributions are $-0.98 \pm 0.06$~\citep{Zhang24_M81} for FRB 20200120E and range from $-0.99$ to $-0.60$~\citep{Ould-Boukattine24} for FRB 20220912A at around 1.4 GHz. Both values suggest relatively flat indices, consistent with the first group.

\section{Discussion}  \label{sec:dis}

\subsection{Bright bursts of various radio sources}

\begin{figure}[!htb]
    \centering
    \includegraphics[width=0.95\linewidth]{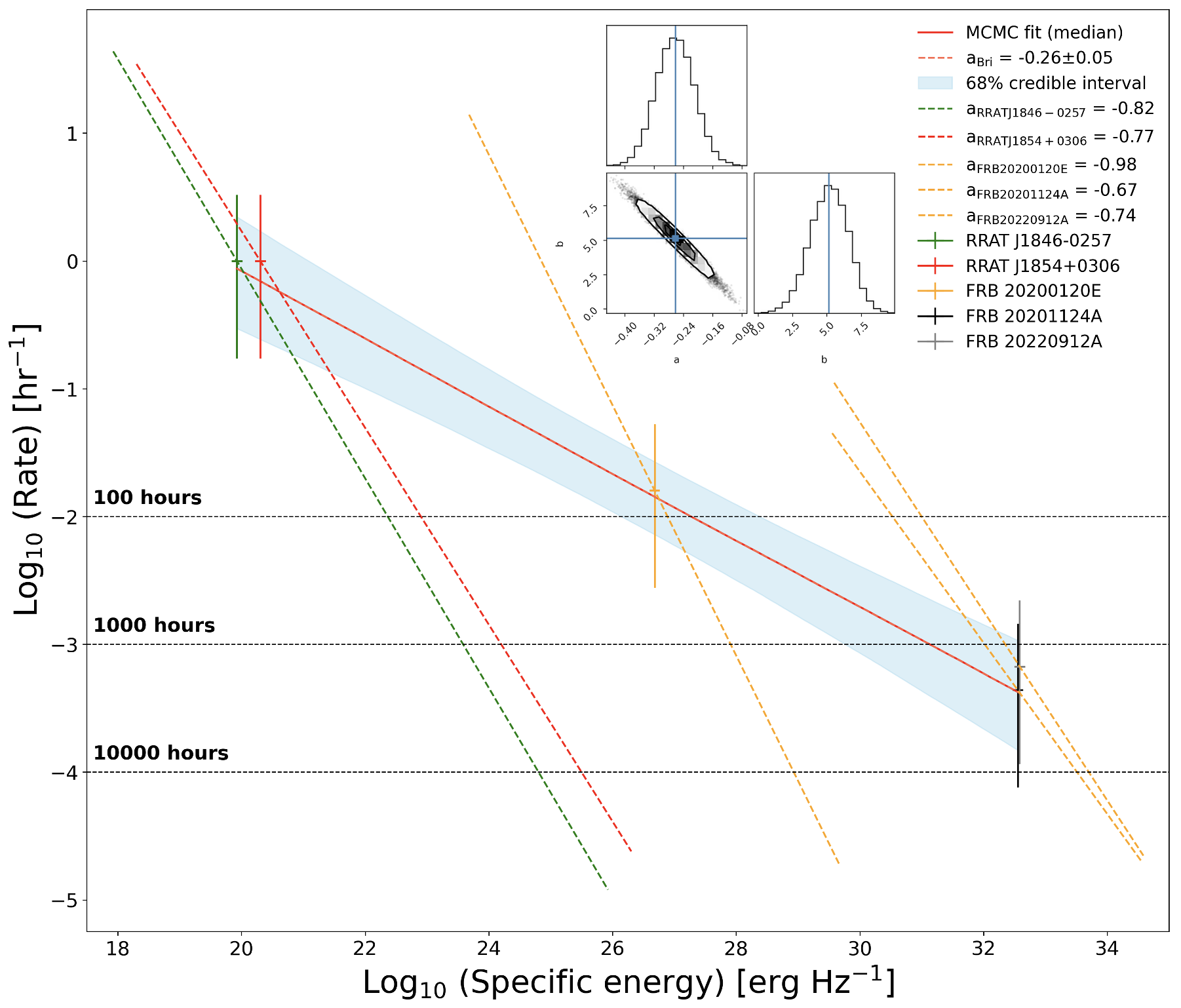}
    \caption{Specific energy of the brightest bursts versus their detection rates for FRB 20201124A, FRB 20220912A, FRB 20200120E and the RRATs~J1846$-$0257 and J1854+0306. The data are fitted with a power law relation (red dashed line) using a Bayesian framework, and the shaded region represents the 68\% credible interval for the fit.
    Posterior distributions and a parameter correlation plot for the
    power law fit parameters are shown in the inset.
    Three horizontal contour lines are added at 100 hours, 1000 hours, and 10000 hours observation time, with their intersections with each source's distribution line indicating the expected rate and energy of the brightest bursts detectable for these observation durations.}
    \label{fig:rate}
\end{figure}

%the brightest bursts with relatively flat indices ($\gtrsim -1$) 
Sources with flat energy distributions ($\alpha \gtrsim -1$) at high energies suggest that their ``bright pulses'' might contribute significantly to the total radio pulse energy release. Therefore, the detection of the brightest bursts from these sources could provide insight into a potential connection between cosmological high-luminosity radio bursts and nearby low-luminosity ones.
%figure2
Figure~\ref{fig:rate} presents the specific energy of the brightest bursts and the corresponding detection rates\footnote{Here, we estimate the rate based on a single detected event, with a $1\sigma$ confidence interval from 0.173 to 3.3~\citep{Gehrels86}.} for the studied sources with flat energy distributions, i.e., RRATs J1846$-$0257 and J1854+0306, FRB 20200120E, FRB 20201124A and FRB 20220912A.
Interestingly, these data points could align with a power-law fit of $\alpha_{\rm Bri} = -0.26 \pm 0.05$.
Due to the sample size of only five points and the stochastic nature of detecting the ``brightest burst'', which usually depends on the telescope sensitivities and monitoring lengths, we will not suggest interpreting this correlation as a physical relationship at this stage.
%Nevertheless, this trend suggests that high-luminosity radio bursts, such as FRBs, could be naturally derived from low-luminosity radio bursts, such as those from pulsars. 
Three horizontal contours at 100, 1000, and 10000 hours mark the expected rate and energy of the brightest detectable bursts at each observation duration. Even extending to 10,000 h, the expected brightest RRAT energies lie below nearby FRBs, and nearby FRBs lie below bright cosmological FRBs. 
This trend suggests that a much flatter index than that observed in current sources such as FRB 20200120E is required to naturally derive high-luminosity radio bursts, such as FRBs, from low-luminosity bursts, such as those from pulsars.
%Nevertheless, this trend suggests that high-luminosity radio bursts, such as FRBs, could be naturally derived from low-luminosity radio bursts, such as those from pulsars. 
%
%Although an extremely shallow power-law index is implied, the flatter energy distributions at higher energies observed in these sources may provide clues for understanding this tentative connection. 

It is worth noting that the power-law indices used here were derived from cumulative distributions, in which the data points are not independent. As a result, our line fitting may produce biased results and underestimate the associated uncertainties. This limitation might be further addressed through more robust methods, such as the maximum likelihood approach introduced by~\citet{James19}, or through future detections of a sufficient number of bright bursts to allow comparison using differential distributions.

%Furthermore, it is notable that if the flat power-law continues (even with an index of $\sim -1$), and given that the number of sources increases with distance$^3$ (neglecting redshift effects, since 1Gpc corresponds to z $\sim 0.23$, which is acceptable for this estimate), a simple order-of-magnitude calculation can be made. For RRATs ($\sim 10^2$ sources, based on the RRATs catalogue, within $\sim$20 kpc according to the ATNF pulsar catalogue) with a typical specific energy of $\sim 10^{20}$erg Hz$^{-1}$, the expected rate for FRB-like bursts with specific energy of $\sim 10^{32}$ erg Hz$^{-1}$ would be $\sim 3 \times 10^5$ day$^{-1}$ Gpc$^{-3}$, indicating that RRATs could indeed produce some FRBs.
%\begin{equation}
%10^2 \,\text{hr}^{-1}\,(20\,\text{kpc})^{-3}
%\times\biggl(\frac{24\,\text{hr}}{\text{day}}\biggr)
%\times\biggl(\frac{\text{Gpc}}{\text{Gpc}}\biggr)^3
%\times\biggl(\frac{10^{20}\,\text{erg}\,\text{Hz}^{-1}}{10^{32}\,\text{erg}\,\text{Hz}^{-1}}\biggr)^{-1}
%\simeq 3\times10^5\,\text{day}^{-1}\,\text{Gpc}^{-3}
%\end{equation}

\subsection{Survey strategies for nearby FRBs} \label{sec:survey}

%%%rewrite the rate derivation of rate
%Previous results are totaly wrong, try my new results
For a telescope operating at a fluence completeness threshold of $F_{\rm det}$, bursts with energy above the threshold $E_{\rm det} = 4\pi D_{\rm det}^2 F_{\rm det}$ are expected to be detected, where $D_{\rm det}$ is the distance to the target source, and assuming a reference burst rate of $R_{\rm ref}$.
%detected burst rate of $R_{\rm det}$.
Applying Equation~\ref{equ:cum_ind}, the burst rate expected for bursts detectable by an arbitrary radio telescope with a fluence threshold $F_{\rm th}$, for a source at a distance of $D_{\rm tar}$, can be written as:
\begin{equation}
\centering
R_{\rm th} \approx R_{\rm ref} \times \left( \frac{F_{\rm th}}{F_{\rm det}} \right)^{\alpha} \left( \frac{D_{\rm tar}}{D_{\rm det}} \right)^{2\alpha}.
\label{equ:rate1}
\end{equation}

As discussed by~\citet{Zhang24_M81}, current FRB surveys may suffer from selection effects, potentially missing FRBs with relatively low luminosity and narrow widths from nearby GC systems.
For example, the current CHIME system only records baseband data for real-time detections with high signal-to-noise ratios (typically between 10 and 12), and at a time resolution of approximately 1\,ms~\citep{CHIME24_baseband}. 
Such a system would only be sensitive to a burst similar to the brightest one from FRB 20200120E (with an effective width of $\le 0.161$\,ms and specific luminosity of $\sim 3.0 \times 10^{30}$ erg\,s$^{-1}$\,Hz$^{-1}$) at distances around $\sim$10\,Mpc, and the event rate is extremely low~\citep{Zhang24_M81}. For more frequent, lower-luminosity bursts from FRB~20200120E, the system is realistically sensitive only to its current distance of $\sim 3$\,Mpc.

%%%previous results 
%A power-law burst energy distribution with index $\beta$ is characterized by $dN/dE \propto E^{\beta}$. For a radio telescope operating at a fluence completeness threshold of $F_{\rm th}$, the expected burst rate for a target source at a distance of $D_{\rm tar}$ can be estimated by~\citep{Kremer23}
%\begin{equation}
%\centering
%R_{\rm th} \approx R_{\rm det} \times \left( \frac{F_{\rm th}}{F_{\rm det}} \right)^{\beta + 1} \left( \frac{D_{\rm tar}}{D_{\rm det}} \right)^{2(\beta + 1)},
%\label{equ:rate1}
%\end{equation}
%where $R_{\rm det}$ is the detected burst rate for a detected source at a distance of $D_{\rm det}$ from an observation with a fluence threshold of $F_{\rm det}$. 

%
\citet{Kremer23} presented prospects for detecting FRBs from GCs in nearby galaxies using 
%FAST~\citep{Jiang20} and MeerKAT~\citep{Bailes20}, 
FAST and MeerKAT, 
suggesting that NGC 4486 (M87) could be the best candidate target.
%
%For relatively large values of $\beta = -2.4$, their predicted FRB detection rates using FAST~\citep{Jiang20} (which has a smaller FOV and higher sensitivity) and MeerKAT~\citep{Bailes20} (which has a larger FOV and lower sensitivity) are $\sim$ 6.75 and 3.31, respectively, FAST would be more efficient.  
%However, these values drop to $\sim$ 0.59 and 1.45 for $\beta = -1.5$ for relatively small values of $\beta = -1.5$, and MeerKAT would perform better. 
%%%new results considering both galaxy radio flux and flatter indices
%galaxy radio flux
However, the fluence threshold ($F_{\rm th}$) of a telescope may be significantly affected by the continuum radio flux densities ($S_{\rm tar}$) from the target galaxies~\citep{Suresh21}, which was not thoroughly discussed in detection rate estimation by~\citet{Kremer23}.
Taking into account $S_{\rm tar}$ and the telescope's corresponding system-equivalent flux density (SEFD), the radiometer equation~\citep{Lorimer04handbook} for estimating the minimum detectable fluence can be written as:
\begin{equation}
F_{\rm min}=\frac{{\rm S/N}_{\rm min} ({\rm SEFD} + S_{\rm tar})}{\sqrt{{\Delta}{\nu}N_p W}} \times W,
\label{equ:limit}
\end{equation}
where we ignore the loss factor, ${\rm S/N}_{\rm min} \sim 7$ is the signal-to-noise ratio (S/N) threshold typically applied in the single-pulse search pipeline, ${\Delta}{\nu}$ is the bandwidth, $N_p$ is the number of polarization, and $W$ is the burst width~\footnote{In this study, we used $W\sim1$\,ms to estimate the sensitivity.}. 
The FAST telescope has a system temperature of $\sim$ 20\,K and a gain of $16 {\rm K/Jy}$~\citep{Jiang20}, resulting in a corresponding SEFD of $S_{\rm FAST} = 1.25$\,Jy,   
while MeerKAT has a system temperature of $\sim$ 18\,K and a gain of $2.8 {\rm K/Jy}$~\citep{Bailes20}, resulting in a corresponding SEFD of $S_{\rm MeerKAT} = 6.43$\,Jy.
%M87 radio flux and et al.
The galaxy M87 could contribute $S_{\rm tar}$ up to 210\,Jy at 1.4\,GHz~\citep{Brown11_M87}, meaning the effective detectable fluence of sources in M87 for FAST and MeerKAT would be approximately 169 and 34 times larger, respectively. 
%FAST, 400MHz(1.05-1.45GHz) or 500MHz(1-1.5GH)? double-check it why?
%CHIME:50K, 1.16K/Jy: ~0.5Jyms, uing 5Jyms??? (why this high, double check it)
%just using the value of Kremer et al. (2023) 
Therefore, for telescopes with high sensitivities, galaxies with relatively small $S_{\rm tar}$, such as NGC 4649 (M60), which has $S_{\rm tar} \sim 0.03$ at 1.4\,GHz~\citep{Brown11_M87} and is estimated to contain approximately 4000 GCs~\citep{Harris13}, could be more efficient search targets.

%%%Parkes CryoPAF
At a given distance, assuming an isotropic source distribution, the number of potential sources is proportional to the field of view (FOV) of the telescope. 
Combining the FOV with Equation~\ref{equ:rate1}, the detection rate of the telescope is given by:
\begin{equation}
\centering
%R \propto F_{\rm th}^{\beta + 1}\cdot D_{\rm tar}^{2(\beta + 1)} \cdot {\rm FOV}.
R \propto F_{\rm th}^{\alpha}\cdot D_{\rm tar}^{2\alpha} \cdot {\rm FOV}.
\label{equ:rate2}
\end{equation}
For smaller values of $\alpha$, the telescope's sensitivity dominates the detection rate, whereas for larger values of $\alpha$ the FOV becomes more important.
Therefore, in addition to telescopes with high sensitivity, such as FAST and MeerKAT, observations with large FOV but relatively lower sensitivity may also be efficient for capturing such events.

\begin{figure}[!htb]
    \centering
    \includegraphics[width=0.95\linewidth]{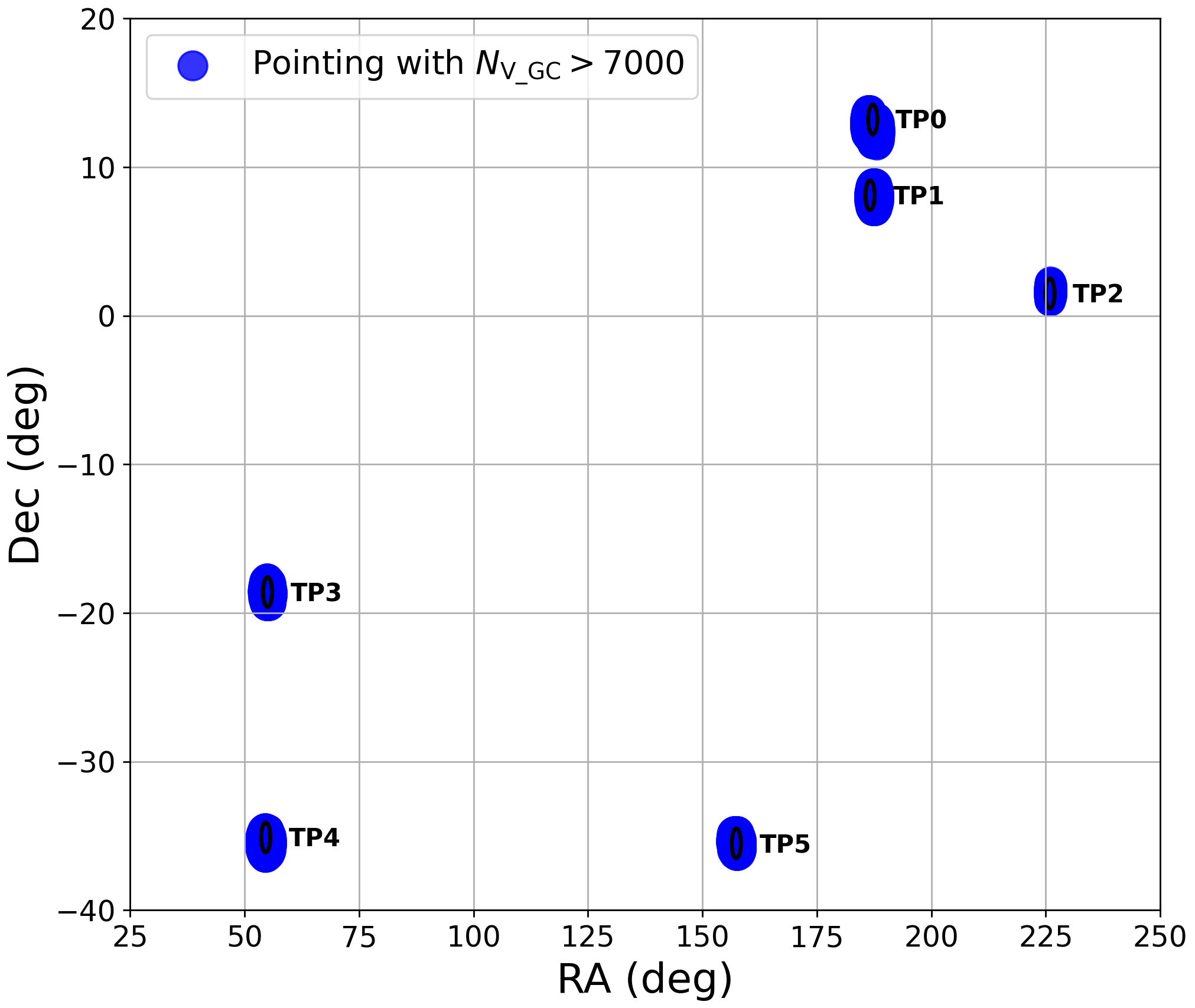}
    \caption{Parkes cryoPAF 1,832 pointings covering more than 7,000 clean GCs, clustered into six groups.
    The sky coverage of the most recommended pointing in each group, TP0 (RA = 187.2 deg; Dec = 13.2 deg), TP1 (RA = 186.6 deg; Dec = 8.1  deg), TP2 (RA = 225.9 deg; Dec = 1.5 deg), TP3 (RA = 55.0 deg; Dec = -18.6 deg), TP4 (RA = 54.6 deg; Dec = -35.1 deg) and TP5 (RA = 157.4 deg; Dec = -35.5 deg), is indicated by black circle.}
    \label{fig:pointing}
\end{figure}

%radio flux of the target sources 
%M87 off-set radio flux
\begin{table*}[!htb]
%\begin{threeparttable}
\begin{center}
\footnotesize
\caption{Properties of the 24 GC systems covered by Parkes cryoPAF pointing TP0 (RA = 187.2 deg; Dec = 13.2 deg).
$d_{\rm L}$ and $D_{\rm TP}$ represent the distance from the GC system centre to the earth and the pointing, respectively.  
References for the galaxy continuum radio flux densities at 1.4\,GHz ($S_{\rm tar, 1.4GHz}$) are: (a)~\citet{Brown11_M87}, (b)~\citet{Allison14}, (c) estimated from National Radio Astronomy Observatory Very Large Array Sky Survey (NVSS, \url{https://www.cv.nrao.edu/nvss/}), (d)~\citet{Nyland17}, (e)~\citet{Oosterloo10}.}
\begin{tabular}{cccccccc}
\hline\hline
Name &  RA (deg) & Dec (deg) & $d_{\rm L}$ (Mpc) & $D_{\rm TP}$ (deg) & $N_{\rm GC}$ & $N_{\rm C,GC}$   & $S_{\rm tar, 1.4GHz}$ (Jy)\\
\hline
NGC4486 (M87)&	187.706&	12.391&	17.0&	0.95&	13000&	6500&	$\sim$ 210 (a)\\
NGC4374 (M84)&	186.265&	12.887&	18.5&	0.96&	4301&	4301&	$\sim$ 6.1 (b)\\
NGC4406 (M86)&	186.549&	12.946&	17.1&	0.68&	2800&	2800&	$<$ 0.1 (c)\\
NGC4473&	187.454&	13.429&	15.2&	0.34&	376&	376&	$<$ 0.1 (a)\\
NGC4435&	186.919&	13.079&	16.6&	0.30&	345&	345&	$<$ 0.1 (c)\\
NGC4459&	187.250&	13.979&	16.0&	0.78&	218&	218&	$<$ 0.1 (d)\\
NGC4474&	187.473&	14.069&	15.5&	0.91&	116&	116&	$<$ 0.1 (c)\\
NGC4458&	187.240&	13.242&	16.3&	0.06&	72&	72&	$<$ 0.1 (e)\\
NGC4387&	186.424&	12.810&	18.0&	0.85&	70&	70&	$<$ 0.1 (e)\\
VCC-940&	186.696&	12.454&	18.7&	0.89&	61&	61&	$<$ 0.1 (c)\\
NGC4479&	187.577&	13.578&	17.4&	0.53&	59&	59&	$<$ 0.1 (c)\\
NGC4478&	187.574&	12.327&	17.0&	0.95&	58&	58&	$<$ 0.1 (c)\\
VCC-1386&	187.964&	12.657&	16.0&	0.92&	27&	27&	$<$ 0.1 (c)\\
%VCC-1386&	187.964&	12.657&	16.1&	0.92&	26&	26&	0.01\\
NGC4476&	187.496&	12.349&	17.7&	0.90&	20&	20&	$<$ 0.1 (c)\\
VCC-871&	186.523&	12.560&	16.0&	0.92&	18&	18&	$<$ 0.1 (c)\\
VCC-965&	186.763&	12.561&	16.9&	0.77&	15&	15&	$<$ 0.1 (c)\\
VCC-1185&	187.348&	12.451&	16.9&	0.76&	14&	14&	$<$ 0.1 (c)\\
NGC4431&	186.864&	12.290&	15.8&	0.97&	11&	11&	$<$ 0.1 (c)\\
VCC-1104&	187.117&	12.824&	15.7&	0.38&	8&	8&	$<$ 0.1 (c)\\
VCC-896&	186.594&	12.787&	16.0&	0.72&	5&	5&	$<$ 0.1 (c)\\
VCC-1105&	187.114&	14.156&	16.0&	0.96&	5&	5&	$<$ 0.1 (c)\\
NGC4486B&	187.633&	12.490&	16.3&	0.83&	4&	4&	$\sim$ 0.4 (c)\\
VCC-1272&	187.567&	13.308&	16.0&	0.37&	3&	3&	$<$ 0.1 (c)\\
VCC-1077&	187.043&	12.807&	16.0&	0.42&	2&	2&	$<$ 0.1 (c)\\
\hline
\end{tabular}
\end{center}
%\end{threeparttable}
\label{tab:TP0}
\end{table*}

\begin{table*}[!htb]
%\begin{threeparttable}
\begin{center}
\footnotesize
\caption{Properties of the GC systems covered by Parkes cryoPAF pointings: 12 systems by TP1 (RA = 186.6 deg; Dec = 8.1 deg), 
3 systems by TP2 (RA = 225.9 deg; Dec = 1.5 deg),
2 systems by TP3 (RA = 55.0 deg; Dec = -18.6 deg),
21 systems by TP4 (RA = 54.6 deg; Dec = -35.1 deg),
and 2 systems by TP5 (RA = 157.4 deg; Dec = -35.5 deg).
References for the galaxy continuum radio flux densities at 1.4\,GHz ($S_{\rm tar, 1.4GHz}$) are: (a)~\citet{White92_M49}, (b)~\citet{Brown11_M87}, (c) estimated from National Radio Astronomy Observatory Very Large Array Sky Survey (NVSS, \url{https://www.cv.nrao.edu/nvss/}), (d)~\citet{Condon98} and 
(e)~\citet{Allison14}.}
\begin{tabular}{l|cccccccc}
\hline\hline
Pointing & Name &  RA (deg) & Dec (deg) & $d_{\rm L}$ (Mpc) & $D_{\rm TP}$ (deg) & $N_{\rm GC}$ & $N_{\rm C,GC}$   & $S_{\rm tar, 1.4GHz}$ (Jy)\\
\hline
TP1 &NGC4472 (M49)&	187.445&	8.000&	17.0&	0.84&	7000&	7000&	$\sim$ 0.8 (a)\\
&NGC4365&	186.118&	7.318&	23.3&	0.92&	3246&	3246&	$<$ 0.1 (b)\\
&NGC4434&	186.903&	8.154&	22.5&	0.30&	141&	141&	$<$ 0.1 (c)\\
&VCC-1254&	187.521&	8.074&	16.0&	0.91&	26&	26&	$<$ 0.1 (c)\\
&VCC-1107&	187.126&	7.325&	16.0&	0.93&	25&	25&	$<$ 0.1 (c)\\
&NGC4464&	187.339&	8.157&	15.8&	0.73&	25&	25&	$<$ 0.1 (c)\\
&NGC4318&	185.680&	8.198&	22.1&	0.92&	18&	18&	$<$ 0.1 (c)\\
&VCC-747&	186.199&	8.992&	18.5&	0.98&	16&	16&	$<$ 0.1 (c)\\
&VCC-1167&	187.301&	7.879&	16.0&	0.73&	14&	14&	$<$ 0.1 (c)\\
&VCC-571&	185.671&	7.950&	23.8&	0.93&	11&	11&	$<$ 0.1 (c)\\
&VCC-1049&	186.978&	8.090&	16.0&	0.37&	8&	8&	$<$ 0.1 (c)\\
&VCC-992&	186.828&	8.213&	16.0&	0.25&	4&	4&	$<$ 0.1 (c)\\
\hline
TP2 &NGC5846&	226.622&	1.606&	24.9&	0.73&	4670&	4670& $<$ 0.1 (b) 	\\  
&NGC5813&	225.297&	1.702&	32.2&	0.64&	2900&	2900& $<$ 0.1 (b)   \\ 
&NGC5845&	226.503&	1.634&	25.9&	0.62&	180&	180&  $<$ 0.1 (c) 	\\ 
\hline
TP3 &NGC1407&	55.049&	-18.580&	28.8&	0.05&	7000&	7000&  $<$ 0.1 (d)	\\  
&NGC1400&	54.879&	-18.688&	26.4&	0.14&	2960&	2960&   $<$ 0.1 (b)	\\  
\hline
TP4 &NGC1399&	54.621&	-35.451&	20.7&	0.35&	6000&	6000& $\sim$ 0.6 (e) \\  
&NGC1404&	54.716&	-35.594&	20.4&	0.50&	770&	770&  $<$ 0.1 (b)	\\  
&NGC1427&	55.581&	-35.393&	20.5&	0.85&	500&	500&  $<$ 0.1 (b)	\\  
&NGC1380&	54.115&	-34.976&	18.9&	0.42&	424&	424&  $<$ 0.1 (b)	\\  
&NGC1387&	54.238&	-35.507&	19.8&	0.50&	390&	390&  $<$ 0.1 (b)	\\  
&NGC1374&	53.819&	-35.226&	19.6&	0.65&	360&	360&  $<$ 0.1 (b)	\\  
&NGC1379&	54.016&	-35.441&	20.0&	0.59&	225&	225&  $<$ 0.1 (b)	\\  
&NGC1380B&	54.287&	-35.195&	22.0&	0.27&	156&	156&  $<$ 0.1 (b)	\\  
&NGC1375&	53.820&	-35.266&	20.0&	0.66&	86&	86&	      $<$ 0.1 (c)    \\  
&NGC1381&	54.132&	-35.295&	18.8&	0.43&	71&	71&	      $<$ 0.1 (c)    \\  
&NGC1380A&	54.198&	-34.740&	16.9&	0.49&	70&	70&       $<$ 0.1 (c) 	\\  
&ESO-358-G2&	53.879&	-34.447&	20.8&	0.88&	60&	60&	      $<$ 0.1 (c)    \\  
&FCC-182&	54.226&	-35.375&	19.6&	0.41&	59&	59&	      $<$ 0.1 (c)    \\  
&NGC1389&	54.299&	-35.746&	20.9&	0.69&	48&	48&	      $<$ 0.1 (c)    \\  
&NGC1428&	55.596&	-35.153&	20.7&	0.82&	42&	42&	      $<$ 0.1 (c)    \\  
&NGC1427A&	55.039&	-35.625&	16.0&	0.63&	38&	38&	      $<$ 0.1 (c)    \\  
&ESO-358-G4&	54.538&	-34.519&	20.7&	0.58&	30&	30&	      $<$ 0.1 (c)    \\  
&FCC-136&	53.623&	-35.546&	19.7&	0.91&	25&	25&	      $<$ 0.1 (c)    \\  
&FCC-254&	55.253&	-35.743&	20.4&	0.83&	6&	6&	      $<$ 0.1 (c)    \\  
&FCC-189&	54.284&	-34.731&	20.0&	0.45&	1&	1&	      $<$ 0.1 (c)    \\  
&FCC-146&	53.798&	-35.323&	20.0&	0.69&	1&	1&	      $<$ 0.1 (c)    \\  
\hline
TP5 &NGC3258&	157.223&	-35.605&	32.1&	0.18&	6000&	6000&  $<$ 0.1 (b)	\\  
&NGC3268&	157.503&	-35.326&	34.8&	0.19&	4750&	4750&  $<$ 0.1 (b)	\\  
\hline
\end{tabular}
\end{center}
%\end{threeparttable}
\label{tab:TP1_5}
\end{table*}

%CryoPAF
A cryogenic phased array receiver (cryoPAF) is currently being completed and tested on the Parkes telescope. The new cryoPAF is expected to have a system temperature of $<$ 20\,K and a gain of $0.735 {\rm K/Jy}$, resulting in a corresponding SEFD of $S_{\rm Parkes} = 27.21$\,Jy. Its wide FOV spans up to 3 square degrees and consists of 72 adjacent beams, covering a frequency range of 700-1950 MHz~\footnote{As cryoPAF is still being tested, the accurate parameters of the full-beam system performance are unavailable for now. We used the values referenced from the Parkes cryoPAF introduction by Alex Dunning: \url{https://research.csiro.au/ratechnologies/wp-content/uploads/sites/295/2022/11/PAFAR2022-Dunning-CryoPAF_for_Parkes.pdf}}. 
%%%Parkes cyroPAF pointings
Based on the cryoPAF’s sky coverage, which is modelled as a circle with a radius of approximately 1 deg in this work, and the catalogue of GC systems~\citep{Harris13} within a relatively close distance ($<$ 40\,Mpc), we conducted a grid analysis of the sky with a 0.1 deg step size to estimate the number of GCs covered by the cryoPAF FOV.  
%M87
Notably, about 20\% of GCs lie within a 4-arcminute radius of the centre of M87~\citep{Strader11}, where the radio background is comparable to $S_{\rm Parkes}$~\citep{Perley17}. Given this, the cryoPAF observation should cover the majority of GCs in M87 with adequate sensitivity, and we adopt a conservative estimate that 50\% of the GCs are clean targets for Parkes observation (denoted as $N_{\rm C,GC}$).
%pointings
As shown in Figure~\ref{fig:pointing}, among the more than 6,000,000 analysed pointings, 1,832 could cover more than 7,000~\footnote{We set this threshold based on the estimated number of $N_{\rm C,GC}$ in GC systems within 40\,Mpc.} clean GCs, clustered into six groups. Based on both the number of $N_{\rm C,GC}$ within the sky coverage and the distance from the pointing to GC system centres, the most recommended pointings for each group are: TP0 (RA = 187.2 deg; Dec = 13.2 deg), TP1 (RA = 186.6 deg; Dec = 8.1  deg), TP2 (RA = 225.9 deg; Dec = 1.5 deg), TP3 (RA = 55.0 deg; Dec = -18.6 deg), TP4 (RA = 54.6 deg; Dec = -35.1 deg) and TP5 (RA = 157.4 deg; Dec = -35.5 deg).
%Parkes_PAF_pointing_above7000.txt  
Full details of all 1,832 pointings are provided in the Supplementary Materials.
%As shown in Figure~\ref{fig:pointing}, among the more than 6,000,000 analysed pointings, 63 could cover more than 10,000 clean GCs, all with Dec near 13 or 8 deg. 
%Pointings TP0 (RA = 187.2 deg; Dec = 13.2 deg) and TP1 (RA = 186.6 deg; Dec = 8.1 deg) cover the largest number of clean GCs in these two Dec ranges. Based on this, we suggest these two pointings as the best targets for searching for FRBs in nearby GCs. 
%
%As listed in Tables~\ref{tab:TP0} and~\ref{tab:TP1}, 
As listed in Tables 1 and 2, Parkes cryoPAF pointing TP0 could cover 24 GC systems, expected to include 21,607 GCs, of which 15,107 are clean targets. TP1 covers 12 GC systems with 10,535 $N_{\rm C,GC}$; TP2 covers 3 GC systems with 7,750 $N_{\rm C,GC}$; TP3 covers 2 GC systems with 9,960 $N_{\rm C,GC}$; TP4 covers 21 GC systems with 9,361 $N_{\rm C,GC}$; and TP5 covers 2 GC systems with 10,750 $N_{\rm C,GC}$.

%flatter indices: -1.5, -2, -2.4, also add: -1, -0.5
\begin{table*}[!htb]
%\begin{threeparttable}
\begin{center}
\footnotesize
\caption{Predicted FRB detection rates in nearby GC systems by FAST, MeerKAT and Parkes cryoPAF.}
\label{tb2}
\begin{tabular}{cccc|ccccc}
\hline\hline
         &            & & & & \multicolumn{2}{c}{FRB detection rate (hr$^{-1}$)}\\
\hline
% Target  &  Instrument & brightest & $\alpha$ = -0.98 & $\alpha$ = -1.5 & $\alpha$ = -2 & $\alpha$ = -2.4 \\
Target  &  Instrument & $N_{\rm GC}$ & $N_{\rm C,GC}$ & $\alpha$ = $-$0.2 & $\alpha$ = $-$0.5 & $\alpha$ = $-$1.0 & $\alpha$ = $-$1.5 & $\alpha$ = $-$2.0 \\
\hline
M60 & FAST & 4000 & 1203 & 0.16(0.01$-$0.48) & 0.35(0.02$-$1.06) & 1.36(0.07$-$4.08) & 5.21(0.27$-$15.64) & 19.98(1.02$-$59.93) \\
M60 & MeerKAT & 4000 & 4000 & 0.37(0.02$-$1.11) & 0.49(0.02$-$1.46) & 0.77(0.04$-$2.31) & 1.21(0.06$-$3.64) & 1.92(0.1$-$5.75) \\
\hline
TP0 & Parkes cryoPAF & 21607 & 15107 & 1.3(0.07$-$3.89) & 1.53(0.08$-$4.6) & 2.03(0.1 $-$6.1)  & 2.72(0.14$-$8.15) & 3.65(0.19$-$10.96) \\
TP1 & Parkes cryoPAF & 10535 & 10535 & 0.88(0.04$-$2.65) & 1.01(0.05$-$3.03) & 1.28(0.07$-$3.84) & 1.65(0.08$-$4.94) & 2.15(0.11$-$6.44) \\
TP2 & Parkes cryoPAF & 7750 & 7750 &0.56(0.03$-$1.68)&0.51(0.03$-$1.54)&0.45(0.02$-$1.34)&0.40(0.02$-$1.19)&0.35(0.02$-$1.06)\\
TP3 & Parkes cryoPAF & 9960 & 9960 
&0.71(0.04$-$2.14)
&0.64(0.03$-$1.92)
&0.54(0.03$-$1.61)
&0.45(0.02$-$1.35)
&0.38(0.02$-$1.13)
\\
TP4 & Parkes cryoPAF & 9361 & 9361
&0.76(0.04$-$2.28)
&0.82(0.04$-$2.47)
&0.94(0.05$-$2.82)
&1.07(0.05$-$3.22)
&1.23(0.06$-$3.69)
\\
TP5 & Parkes cryoPAF & 10750 & 10750       
&0.72(0.04$-$2.16)
&0.58(0.03$-$1.75)
&0.41(0.02$-$1.24)
&0.29(0.01$-$0.87)
&0.21(0.01$-$0.62)
\\
%TP & cryoPAF & 0.38(0.02$-$1.13) & 2.79(0.14$-$8.28) & 8.03(0.40$-$23.86) & 22.21(1.12$-$65.97) & 50.11(2.52$-$148.82) \\
%M87 & FAST & - & - & 0.59(0.03$-$1.78) & 2.29(0.12$-$6.87) & 6.75(0.34$-$20.26) \\
%M87 & MeerKat  & - & - & 1.45(0.07$-$4.36) & 2.29(0.12$-$6.88) & 3.31(0.17$-$9.92) \\
\hline
\end{tabular}
\end{center}
%\end{threeparttable}
\label{tab:rate}
\end{table*}

%the predict event rate 
Furthermore, we estimated the predicted FRB detection rates for FAST and MeerKAT targeting M60, as well as for Parkes cryoPAF targeting TP0 and TP1, using power-law indices $\alpha = -0.2, \,-0.5, \,-1.0, \,-1.5$ and $-2.0$. 
%same parameters
Most of the essential values used in this calculation are similar to~\citet{Kremer23}.
The fluence thresholds of FAST and MeerKAT are 0.015\,Jy\,ms~\citep{Niu21} and 0.09\,Jy\,ms~\citep{Bailes20}. 
The reference values for the reference rate ($R_{\rm ref}\sim0.07$\,hr$^{-1}$), fluence threshold ($F_{\rm det}\sim5$\,Jy\,ms), and source distance ( $D_{\rm det}\sim3.63$\,Mpc) were derived from CHIME’s observations of FRB 20200120E~\citep{CHIME21, Bhardwaj21, Kirsten22}. The detection of a single FRB repeater by CHIME in the GCs within the M81 region suggests a specific abundance of $N_{\rm obs}\sim1/900\approx1.1\times 10^{-3}$~\citep{Bhardwaj21,Harris13,Kremer23}, with a 90\% confidence range of $(5.6\times 10^{-3},\, 3.3\times 10^{-3})$, incorporating the Poisson probability~\citep{Lu22}.   
%our parameters
For GC systems with $S_{\rm tar, 1.4GHz} < 0.1$\,Jy, which are negligible compared to the SEFDs of all three telescopes, we ignored their effect on the minimum detectable fluence. For GCs with larger $S_{\rm tar, 1.4GHz}$, we used the effective minimum detectable fluence calculated using Equation~\ref{equ:limit}.

%%%discuss the prediction
%As presented in Table~\ref{tab:rate}, 
As presented in Table 3, 
for steep power-law indices ($\alpha = -1.5, \,-2.0$), FAST is the most powerful telescope for detecting nearby FRBs, with a detection rate of up to 20\,hr$^{-1}$. Even at the lower detection rate limits, only about 3 hours of observation are required to expect a new detection.
In contrast, Parkes cryoPAF is more efficient for flatter indices ($\alpha = -0.2, \,-0.5, \,-1.5$), with 10 to 20 hours of observation typically required to ensure detecting a new source.

%
%Thus, the survey strategies will differ significantly for sources with different energy distributions. If the burst distribution at high energy is extremely flat, surveys with large FOV but relatively low-sensitivity, such as the forthcoming Cryogenic Phased Array Feed (CryoPAF) receiver for the Parkes telescope, or the proposed compact all-sky phased array~\citep{Luo24}, could efficiently capture such events.

%\subsection{Models?}

\section{Conclusion}   \label{sec:con}

%history
%Before the discovery of a population of FRBs~\citep{Thornton13} that supported the astrophysical origin of the ``Lorimer burst''~\citep{Lorimer07}, pulsars~\citep{Manchester05} were considered the primary sources of bright, short-duration ($\lesssim $ second) radio pulses, which require coherent emission~\citep{Pietka15}. While FRBs also necessitate coherent emission, their luminosities are more than ten orders of magnitude higher. Initially, the apparent one-off nature of FRBs led to favouring catastrophic scenarios~\citep{Zhang23_FRBModel}, distinguishing them from pulsars.
%
%However, the discovery of repeating FRBs~\citep{Spitler16}, and the investigation that some Galactic radio pulsing sources might be one-off~\citep{Keane16,Zhang24_Parkes}, have reduced this distinction. 
%
%Further observations of fainter FRB bursts~\citep{Bochenek20, CHIME20, Nimmo23, Zhang24_M81} and brighter pulses from magnetars or pulsars~\citep{Kirsten21, Good2020ATel, Bera19} suggest no clear gap between the isotropic-equivalent energy releases of various radio pulses~\citep{Zhang24_M81}.
%
%Additionally, polarimetric similarities between FRBs and pulsars have been observed~\citep{Feng22,Niu24}.
%
%These findings highlight the importance of exploring potential connections between FRBs and pulsars, at least for certain sources.

%our work
In this study, we have presented similarities in the high-energy distributions of two cosmological repeating FRBs (FRB 20201124A and FRB 20220912A), one nearby FRB 20200120E, and two pulsars (RRATs J1846$-$0257 and J1854+0306). 
All these sources exhibit power-law indices of $\alpha \gtrsim -1$, suggesting that their ``bright pulses'' might significantly contribute to the total radio pulse energy release. 
Furthermore, the brightest bursts from these sources could be fitted with a power-law model of $\alpha_{\rm Bri} = -0.26 \pm 0.05$. 
Although we will not suggest interpreting this correlation as a physical relationship, the trend suggests that an extremely flat index is required to naturally derive high-luminosity radio bursts, such as FRBs, from low-luminosity bursts, such as those from pulsars.
%However, considering the limited sample size, we advise caution in interpreting this correlation as physical.
%
%Notably, the two pulsars studied have relatively high surface magnetic fields ($\sim 2 \times 10^{13}$ G), which may be related to FRB models involving magnetars~\citep{Zhang23_FRBModel}, known for their strong magnetic fields~\citep{Esposito21}.
%
We encourage further observations to expand the sample of high-energy bursts from both FRB-like and pulsar-like sources. 

%
%%searching for nearby FRBs
%talk about more for CHIME survey, why no nearby FRB > 10Mpc were detected.
%
As discussed in Section~\ref{sec:survey}, current FRB surveys may miss FRBs with relatively low luminosity and narrow widths from nearby GC systems.
%As discussed by~\citet{Zhang24_M81}, current FRB surveys may suffer from selection effects, potentially missing FRBs with relatively low luminosity and narrow widths from nearby GC systems.For example, the current CHIME system only records baseband data for real-time detections with high signal-to-noise ratios (typically between 10 and 12), and at a time resolution of approximately 1\,ms~\citep{CHIME24_baseband}. Such a system would only be sensitive to a burst similar to the brightest one from FRB 20200120E (with an effective width of $\le 0.161$\,ms and specific luminosity of $\sim 3.0 \times 10^{30}$ erg\,s$^{-1}$\,Hz$^{-1}$) at distances around $\sim$10\,Mpc, and the event rate is extremely low~\citep{Zhang24_M81}. For more frequent, lower-luminosity bursts from FRB~20200120E, the system is realistically sensitive only to its current distance of $\sim 3$\,Mpc.
%Current FRB surveys may be limited by selection effects, potentially missing FRBs with relative low-luminosity from the nearby GC systems~\citep{Zhang24_M81}.
%
Survey strategies will need to vary to search for sources with different energy distributions. 
%future observation based on the Survey strategie
In this work, we have discussed detailed survey strategies for FAST, MeerKAT and Parkes cryoPAF for FRB searches in nearby GCs and suggested optimal survey targets.
%constrain
In principle, conducting combined observations for these targets using FAST for about $3$ hours and Parkes cryoPAF for $10-20$ hours could potentially lead to the discovery of new FRBs in nearby GCs, in practice.  
Additionally, a lack of detection after even a few hours of FAST observation can still provide strict constraints on the combination of $\alpha$ values and intrinsic activity rates, or FRB rates.

%\clearpage
%% Also note that the akcnowlodgment environment does not support long amounts of text. If you have a lot of people and institutions to acknowledge, do not use this command. Instead, create a new 
\section*{Acknowledgments}
We thank the reviewer for the valuable and detailed comments, which have greatly improved the quality of this manuscript.
This research has been partially funded by the International Partnership Program of Chinese Academy of Sciences for Grand Challenges (114332KYSB20210018), the National SKA Program of China (2022SKA0130100), the National Natural Science Foundation of China (grant Nos. 12041306,12273113,12233002,12003028,12321003), the CAS Project for Young Scientists in Basic Research (Grant No. YSBR-063), , and the ACAMAR Postdoctoral Fellow. \\

\section*{Supplementary Materials}
Full details of all 1,832 Parkes cryoPAF pointings, which cover more than 7,000 clean GCs, are provided in the file ``Parkes\_PAF\_pointing\_above7000.txt''. 


\clearpage
\renewcommand{\baselinestretch}{0.8}
\bibliography{ms}{}

\begin{thebibliography}{}
\expandafter\ifx\csname natexlab\endcsname\relax\def\natexlab#1{#1}\fi
\providecommand{\url}[1]{\href{#1}{#1}}
\providecommand{\dodoi}[1]{doi:~\href{http://doi.org/#1}{\nolinkurl{#1}}}
\providecommand{\doeprint}[1]{\href{http://ascl.net/#1}{\nolinkurl{http://ascl.net/#1}}}
\providecommand{\doarXiv}[1]{\href{https://arxiv.org/abs/#1}{\nolinkurl{https://arxiv.org/abs/#1}}}

\bibitem[{{Allison} {et~al.}(2014){Allison}, {Sadler}, \& {Meekin}}]{Allison14}
{Allison}, J.~R., {Sadler}, E.~M., \& {Meekin}, A.~M. 2014, \mnras, 440, 696,
  \dodoi{10.1093/mnras/stu289}

\bibitem[{{Bailes} {et~al.}(2020){Bailes}, {Jameson}, {Abbate}, {Barr}, {Bhat},
  {Bondonneau}, {Burgay}, {Buchner}, {Camilo}, {Champion}, {Cognard},
  {Demorest}, {Freire}, {Gautam}, {Geyer}, {Griessmeier}, {Guillemot}, {Hu},
  {Jankowski}, {Johnston}, {Karastergiou}, {Karuppusamy}, {Kaur}, {Keith},
  {Kramer}, {van Leeuwen}, {Lower}, {Maan}, {McLaughlin}, {Meyers},
  {Os{\l}owski}, {Oswald}, {Parthasarathy}, {Pennucci}, {Posselt}, {Possenti},
  {Ransom}, {Reardon}, {Ridolfi}, {Schollar}, {Serylak}, {Shaifullah},
  {Shamohammadi}, {Shannon}, {Sobey}, {Song}, {Spiewak}, {Stairs}, {Stappers},
  {van Straten}, {Szary}, {Theureau}, {Venkatraman Krishnan}, {Weltevrede},
  {Wex}, {Abbott}, {Adams}, {Burger}, {Gamatham}, {Gouws}, {Horn}, {Hugo},
  {Joubert}, {Manley}, {McAlpine}, {Passmoor}, {Peens-Hough}, {Ramudzuli},
  {Rust}, {Salie}, {Schwardt}, {Siebrits}, {Van Tonder}, {Van Tonder}, \&
  {Welz}}]{Bailes20}
{Bailes}, M., {Jameson}, A., {Abbate}, F., {et~al.} 2020, \pasa, 37, e028,
  \dodoi{10.1017/pasa.2020.19}

\bibitem[{{Bera} \& {Chengalur}(2019)}]{Bera19}
{Bera}, A., \& {Chengalur}, J.~N. 2019, \mnras, 490, L12,
  \dodoi{10.1093/mnrasl/slz140}

\bibitem[{{Bhardwaj} {et~al.}(2021){Bhardwaj}, {Gaensler}, {Kaspi},
  {Landecker}, {Mckinven}, {Michilli}, {Pleunis}, {Tendulkar}, {Andersen},
  {Boyle}, {Cassanelli}, {Chawla}, {Cook}, {Dobbs}, {Fonseca}, {Kaczmarek},
  {Leung}, {Masui}, {Mnchmeyer}, {Ng}, {Rafiei-Ravandi}, {Scholz}, {Shin},
  {Smith}, {Stairs}, \& {Zwaniga}}]{Bhardwaj21}
{Bhardwaj}, M., {Gaensler}, B.~M., {Kaspi}, V.~M., {et~al.} 2021, \apjl, 910,
  L18, \dodoi{10.3847/2041-8213/abeaa6}

\bibitem[{{Bochenek} {et~al.}(2020){Bochenek}, {Ravi}, {Belov}, {Hallinan},
  {Kocz}, {Kulkarni}, \& {McKenna}}]{Bochenek20}
{Bochenek}, C.~D., {Ravi}, V., {Belov}, K.~V., {et~al.} 2020, \nat, 587, 59,
  \dodoi{10.1038/s41586-020-2872-x}

\bibitem[{{Brown} {et~al.}(2011){Brown}, {Jannuzi}, {Floyd}, \&
  {Mould}}]{Brown11_M87}
{Brown}, M. J.~I., {Jannuzi}, B.~T., {Floyd}, D. J.~E., \& {Mould}, J.~R. 2011,
  \apjl, 731, L41, \dodoi{10.1088/2041-8205/731/2/L41}

\bibitem[{{Burke-Spolaor} {et~al.}(2012){Burke-Spolaor}, {Johnston}, {Bailes},
  {Bates}, {Bhat}, {Burgay}, {Champion}, {D'Amico}, {Keith}, {Kramer}, {Levin},
  {Milia}, {Possenti}, {Stappers}, \& {van Straten}}]{Burke-Spolaor12}
{Burke-Spolaor}, S., {Johnston}, S., {Bailes}, M., {et~al.} 2012, \mnras, 423,
  1351, \dodoi{10.1111/j.1365-2966.2012.20998.x}

\bibitem[{{CHIME/FRB Collaboration} {et~al.}(2020){CHIME/FRB Collaboration},
  {Andersen}, {Bandura}, {Bhardwaj}, {Bij}, {Boyce}, {Boyle}, {Brar},
  {Cassanelli}, {Chawla}, {Chen}, {Cliche}, {Cook}, {Cubranic}, {Curtin},
  {Denman}, {Dobbs}, {Dong}, {Fandino}, {Fonseca}, {Gaensler}, {Giri}, {Good},
  {Halpern}, {Hill}, {Hinshaw}, {H{\"o}fer}, {Josephy}, {Kania}, {Kaspi},
  {Landecker}, {Leung}, {Li}, {Lin}, {Masui}, {McKinven}, {Mena-Parra},
  {Merryfield}, {Meyers}, {Michilli}, {Milutinovic}, {Mirhosseini},
  {M{\"u}nchmeyer}, {Naidu}, {Newburgh}, {Ng}, {Patel}, {Pen},
  {Pinsonneault-Marotte}, {Pleunis}, {Quine}, {Rafiei-Ravandi}, {Rahman},
  {Ransom}, {Renard}, {Sanghavi}, {Scholz}, {Shaw}, {Shin}, {Siegel}, {Singh},
  {Smegal}, {Smith}, {Stairs}, {Tan}, {Tendulkar}, {Tretyakov}, {Vanderlinde},
  {Wang}, {Wulf}, \& {Zwaniga}}]{CHIME20}
{CHIME/FRB Collaboration}, {Andersen}, B.~C., {Bandura}, K.~M., {et~al.} 2020,
  \nat, 587, 54, \dodoi{10.1038/s41586-020-2863-y}

\bibitem[{{Chime/Frb Collaboration} {et~al.}(2020){Chime/Frb Collaboration},
  {Amiri}, {Andersen}, {Bandura}, {Bhardwaj}, {Boyle}, {Brar}, {Chawla},
  {Chen}, {Cliche}, {Cubranic}, {Deng}, {Denman}, {Dobbs}, {Dong}, {Fandino},
  {Fonseca}, {Gaensler}, {Giri}, {Good}, {Halpern}, {Hessels}, {Hill},
  {H{\"o}fer}, {Josephy}, {Kania}, {Karuppusamy}, {Kaspi}, {Keimpema},
  {Kirsten}, {Landecker}, {Lang}, {Leung}, {Li}, {Lin}, {Marcote}, {Masui},
  {McKinven}, {Mena-Parra}, {Merryfield}, {Michilli}, {Milutinovic},
  {Mirhosseini}, {Naidu}, {Newburgh}, {Ng}, {Nimmo}, {Paragi}, {Patel}, {Pen},
  {Pinsonneault-Marotte}, {Pleunis}, {Rafiei-Ravandi}, {Rahman}, {Ransom},
  {Renard}, {Sanghavi}, {Scholz}, {Shaw}, {Shin}, {Siegel}, {Singh}, {Smegal},
  {Smith}, {Stairs}, {Tendulkar}, {Tretyakov}, {Vanderlinde}, {Wang}, {Wang},
  {Wulf}, {Yadav}, \& {Zwaniga}}]{Chime20_Li}
{Chime/Frb Collaboration}, {Amiri}, M., {Andersen}, B.~C., {et~al.} 2020, \nat,
  582, 351, \dodoi{10.1038/s41586-020-2398-2}

\bibitem[{{CHIME/FRB Collaboration} {et~al.}(2021){CHIME/FRB Collaboration},
  {Amiri}, {Andersen}, {Bandura}, {Berger}, {Bhardwaj}, {Boyce}, {Boyle},
  {Brar}, {Breitman}, {Cassanelli}, {Chawla}, {Chen}, {Cliche}, {Cook},
  {Cubranic}, {Curtin}, {Deng}, {Dobbs}, {Dong}, {Eadie}, {Fandino}, {Fonseca},
  {Gaensler}, {Giri}, {Good}, {Halpern}, {Hill}, {Hinshaw}, {Josephy},
  {Kaczmarek}, {Kader}, {Kania}, {Kaspi}, {Landecker}, {Lang}, {Leung}, {Li},
  {Lin}, {Masui}, {McKinven}, {Mena-Parra}, {Merryfield}, {Meyers}, {Michilli},
  {Milutinovic}, {Mirhosseini}, {M{\"u}nchmeyer}, {Naidu}, {Newburgh}, {Ng},
  {Patel}, {Pen}, {Petroff}, {Pinsonneault-Marotte}, {Pleunis},
  {Rafiei-Ravandi}, {Rahman}, {Ransom}, {Renard}, {Sanghavi}, {Scholz}, {Shaw},
  {Shin}, {Siegel}, {Sikora}, {Singh}, {Smith}, {Stairs}, {Tan}, {Tendulkar},
  {Vanderlinde}, {Wang}, {Wulf}, \& {Zwaniga}}]{CHIME21}
{CHIME/FRB Collaboration}, {Amiri}, M., {Andersen}, B.~C., {et~al.} 2021,
  \apjs, 257, 59, \dodoi{10.3847/1538-4365/ac33ab}

\bibitem[{{Chime/Frb Collaboration} {et~al.}(2023){Chime/Frb Collaboration},
  {Andersen}, {Bandura}, {Bhardwaj}, {Boyle}, {Brar}, {Cassanelli},
  {Chatterjee}, {Chawla}, {Cook}, {Curtin}, {Dobbs}, {Dong}, {Faber},
  {Fandino}, {Fonseca}, {Gaensler}, {Giri}, {Herrera-Martin}, {Hill}, {Ibik},
  {Josephy}, {Kaczmarek}, {Kader}, {Kaspi}, {Landecker}, {Lanman}, {Lazda},
  {Leung}, {Lin}, {Masui}, {McKinven}, {Mena-Parra}, {Meyers}, {Michilli},
  {Ng}, {Pandhi}, {Pearlman}, {Pen}, {Petroff}, {Pleunis}, {Rafiei-Ravandi},
  {Rahman}, {Ransom}, {Renard}, {Sand}, {Sanghavi}, {Scholz}, {Shah}, {Shin},
  {Siegel}, {Smith}, {Stairs}, {Su}, {Tendulkar}, {Vanderlinde}, {Wang},
  {Wulf}, \& {Zwaniga}}]{CHIME23}
{Chime/Frb Collaboration}, {Andersen}, B.~C., {Bandura}, K., {et~al.} 2023,
  \apj, 947, 83, \dodoi{10.3847/1538-4357/acc6c1}

\bibitem[{{CHIME/FRB Collaboration} {et~al.}(2024){CHIME/FRB Collaboration},
  {Amiri}, {Andersen}, {Andrew}, {Bandura}, {Bhardwaj}, {Boyle}, {Brar},
  {Breitman}, {Cassanelli}, {Chawla}, {Cook}, {Curtin}, {Dobbs}, {Dong},
  {Eadie}, {Fonseca}, {Gaensler}, {Giri}, {Herrera-Martin}, {Hopkins}, {Ibik},
  {Joseph}, {Kaczmarek}, {Kader}, {Kaspi}, {Lanman}, {Lazda}, {Leung}, {Liu},
  {Masui}, {McKinven}, {Mena-Parra}, {Merryfield}, {Michilli}, {Ng}, {Nimmo},
  {Noble}, {Pandhi}, {Patel}, {Pearlman}, {Pen}, {Petroff}, {Pleunis},
  {Rafiei-Ravandi}, {Rahman}, {Ransom}, {Sand}, {Scholz}, {Shah}, {Shin},
  {Shpunarska}, {Siegel}, {Smith}, {Stairs}, {Stenning}, {Vanderlinde}, {Wang},
  {White}, \& {Wulf}}]{CHIME24_baseband}
{CHIME/FRB Collaboration}, {Amiri}, M., {Andersen}, B.~C., {et~al.} 2024, \apj,
  969, 145, \dodoi{10.3847/1538-4357/ad464b}

\bibitem[{{Chime/Frb Collaboration} {et~al.}(2022){Chime/Frb Collaboration},
  {Bandura}, {Bhardwaj}, {Boyle}, {Brar}, {Breitman}, {Cassanelli},
  {Chatterjee}, {Chawla}, {Cliche}, {Cubranic}, {Curtin}, {Deng}, {Dobbs},
  {Dong}, {Fonseca}, {Gaensler}, {Giri}, {Good}, {Hill}, {Josephy},
  {Kaczmarek}, {Kader}, {Kania}, {Kaspi}, {Leung}, {Li}, {Lin}, {Masui},
  {McKinven}, {Mena-Parra}, {Merryfield}, {Meyers}, {Michilli}, {Naidu},
  {Newburgh}, {Ng}, {Ordog}, {Patel}, {Pearlman}, {Pen}, {Petroff}, {Pleunis},
  {Rafiei-Ravandi}, {Rahman}, {Ransom}, {Renard}, {Sanghavi}, {Scholz}, {Shaw},
  {Shin}, {Siegel}, {Singh}, {Smith}, {Stairs}, {Tan}, {Tendulkar},
  {Vanderlinde}, {Wiebe}, {Wulf}, \& {Zwaniga}}]{Chime22_subs}
{Chime/Frb Collaboration}, Andersen, B.~C., {Bandura}, K., {Bhardwaj}, M.,
  {et~al.} 2022, \nat, 607, 256, \dodoi{10.1038/s41586-022-04841-8}

\bibitem[{{Condon} {et~al.}(1998){Condon}, {Cotton}, {Greisen}, {Yin},
  {Perley}, {Taylor}, \& {Broderick}}]{Condon98}
{Condon}, J.~J., {Cotton}, W.~D., {Greisen}, E.~W., {et~al.} 1998, \aj, 115,
  1693, \dodoi{10.1086/300337}

\bibitem[{{Cruces} {et~al.}(2021){Cruces}, {Spitler}, {Scholz}, {Lynch},
  {Seymour}, {Hessels}, {Gouiff{\'e}s}, {Hilmarsson}, {Kramer}, \&
  {Munjal}}]{Cruces21}
{Cruces}, M., {Spitler}, L.~G., {Scholz}, P., {et~al.} 2021, \mnras, 500, 448,
  \dodoi{10.1093/mnras/staa3223}

\bibitem[{{Foreman-Mackey} {et~al.}(2013){Foreman-Mackey}, {Hogg}, {Lang}, \&
  {Goodman}}]{emcee2013}
{Foreman-Mackey}, D., {Hogg}, D.~W., {Lang}, D., \& {Goodman}, J. 2013, \pasp,
  125, 306, \dodoi{10.1086/670067}

\bibitem[{{Gehrels}(1986)}]{Gehrels86}
{Gehrels}, N. 1986, \apj, 303, 336, \dodoi{10.1086/164079}

\bibitem[{{Geyer} {et~al.}(2021){Geyer}, {Serylak}, {Abbate}, {Bailes},
  {Buchner}, {Chilufya}, {Johnston}, {Karastergiou}, {Main}, {van Straten}, \&
  {Shamohammadi}}]{Geyer21}
{Geyer}, M., {Serylak}, M., {Abbate}, F., {et~al.} 2021, \mnras, 505, 4468,
  \dodoi{10.1093/mnras/stab1501}

\bibitem[{{Good} \& {Chime/Frb Collaboration}(2020)}]{Good2020ATel}
{Good}, D., \& {Chime/Frb Collaboration}. 2020, The Astronomer's Telegram,
  14074, 1

\bibitem[{{Harris} {et~al.}(2013){Harris}, {Harris}, \& {Alessi}}]{Harris13}
{Harris}, W.~E., {Harris}, G. L.~H., \& {Alessi}, M. 2013, \apj, 772, 82,
  \dodoi{10.1088/0004-637X/772/2/82}

\bibitem[{{Hewitt} {et~al.}(2022){Hewitt}, {Snelders}, {Hessels}, {Nimmo},
  {Jahns}, {Spitler}, {Gourdji}, {Hilmarsson}, {Michilli}, {Ould-Boukattine},
  {Scholz}, \& {Seymour}}]{Hewitt22}
{Hewitt}, D.~M., {Snelders}, M.~P., {Hessels}, J.~W.~T., {et~al.} 2022, \mnras,
  515, 3577, \dodoi{10.1093/mnras/stac1960}

\bibitem[{{Huang} {et~al.}(2025){Huang}, {Zhang}, {Xu}, {Hao}, {Lee}, {Zhang},
  {Wang}, {Cao}, {Zhou}, {Xu}, {Li}, {Xu}, {Wang}, {Jiang}, {Guo}, {Xue},
  {Shen}, {Wang}, {Men}, {Chen}, {Wu}, \& {Wang}}]{Huang25}
{Huang}, Y.-X., {Zhang}, J.-S., {Xu}, H., {et~al.} 2025, arXiv e-prints,
  arXiv:2504.03569, \dodoi{10.48550/arXiv.2504.03569}

\bibitem[{{Jahns} {et~al.}(2023){Jahns}, {Spitler}, {Nimmo}, {Hewitt},
  {Snelders}, {Seymour}, {Hessels}, {Gourdji}, {Michilli}, \&
  {Hilmarsson}}]{Jahns23}
{Jahns}, J.~N., {Spitler}, L.~G., {Nimmo}, K., {et~al.} 2023, \mnras, 519, 666,
  \dodoi{10.1093/mnras/stac3446}

\bibitem[{{James} {et~al.}(2019){James}, {Ekers}, {Macquart}, {Bannister}, \&
  {Shannon}}]{James19}
{James}, C.~W., {Ekers}, R.~D., {Macquart}, J.~P., {Bannister}, K.~W., \&
  {Shannon}, R.~M. 2019, \mnras, 483, 1342, \dodoi{10.1093/mnras/sty3031}

\bibitem[{{Jiang} {et~al.}(2020){Jiang}, {Tang}, {Hou}, {Liu}, {Kr{\v{c}}o},
  {Qian}, {Sun}, {Ching}, {Liu}, {Duan}, {Yue}, {Gan}, {Yao}, {Li}, {Pan},
  {Yu}, {Liu}, {Li}, {Peng}, {Yan}, \& {FAST Collaboration}}]{Jiang20}
{Jiang}, P., {Tang}, N.-Y., {Hou}, L.-G., {et~al.} 2020, Research in Astronomy
  and Astrophysics, 20, 064, \dodoi{10.1088/1674-4527/20/5/64}

\bibitem[{{Kirsten} {et~al.}(2021){Kirsten}, {Snelders}, {Jenkins}, {Nimmo},
  {van den Eijnden}, {Hessels}, {Gawro{\'n}ski}, \& {Yang}}]{Kirsten21}
{Kirsten}, F., {Snelders}, M.~P., {Jenkins}, M., {et~al.} 2021, Nature
  Astronomy, 5, 414, \dodoi{10.1038/s41550-020-01246-3}

\bibitem[{{Kirsten} {et~al.}(2022){Kirsten}, {Marcote}, {Nimmo}, {Hessels},
  {Bhardwaj}, {Tendulkar}, {Keimpema}, {Yang}, {Snelders}, {Scholz},
  {Pearlman}, {Law}, {Peters}, {Giroletti}, {Paragi}, {Bassa}, {Hewitt},
  {Bach}, {Bezrukovs}, {Burgay}, {Buttaccio}, {Conway}, {Corongiu}, {Feiler},
  {Forss{\'e}n}, {Gawro{\'n}ski}, {Karuppusamy}, {Kharinov}, {Lindqvist},
  {Maccaferri}, {Melnikov}, {Ould-Boukattine}, {Possenti}, {Surcis}, {Wang},
  {Yuan}, {Aggarwal}, {Anna-Thomas}, {Bower}, {Blaauw}, {Burke-Spolaor},
  {Cassanelli}, {Clarke}, {Fonseca}, {Gaensler}, {Gopinath}, {Kaspi}, {Kassim},
  {Lazio}, {Leung}, {Li}, {Lin}, {Masui}, {Mckinven}, {Michilli}, {Mikhailov},
  {Ng}, {Orbidans}, {Pen}, {Petroff}, {Rahman}, {Ransom}, {Shin}, {Smith},
  {Stairs}, \& {Vlemmings}}]{Kirsten22}
{Kirsten}, F., {Marcote}, B., {Nimmo}, K., {et~al.} 2022, \nat, 602, 585,
  \dodoi{10.1038/s41586-021-04354-w}

\bibitem[{{Kirsten} {et~al.}(2024){Kirsten}, {Ould-Boukattine}, {Herrmann},
  {Gawro{\'n}ski}, {Hessels}, {Lu}, {Snelders}, {Chawla}, {Yang}, {Blaauw},
  {Nimmo}, {Puchalska}, {Wolak}, \& {van Ruiten}}]{Kirsten23}
{Kirsten}, F., {Ould-Boukattine}, O.~S., {Herrmann}, W., {et~al.} 2024, Nature
  Astronomy, 8, 337, \dodoi{10.1038/s41550-023-02153-z}

\bibitem[{{Kremer} {et~al.}(2023){Kremer}, {Li}, {Lu}, {Piro}, \&
  {Zhang}}]{Kremer23}
{Kremer}, K., {Li}, D., {Lu}, W., {Piro}, A.~L., \& {Zhang}, B. 2023, \apj,
  944, 6, \dodoi{10.3847/1538-4357/acabbf}

\bibitem[{{Kuzmin}(2007)}]{Kuzmin07}
{Kuzmin}, A.~D. 2007, \apss, 308, 563, \dodoi{10.1007/s10509-007-9347-5}

\bibitem[{{Li} {et~al.}(2021){Li}, {Wang}, {Zhu}, {Zhang}, {Zhang}, {Duan},
  {Zhang}, {Feng}, {Tang}, {Chatterjee}, {Cordes}, {Cruces}, {Dai}, {Gajjar},
  {Hobbs}, {Jin}, {Kramer}, {Lorimer}, {Miao}, {Niu}, {Niu}, {Pan}, {Qian},
  {Spitler}, {Werthimer}, {Zhang}, {Wang}, {Xie}, {Yue}, {Zhang}, {Zhi}, \&
  {Zhu}}]{Li21}
{Li}, D., {Wang}, P., {Zhu}, W.~W., {et~al.} 2021, \nat, 598, 267,
  \dodoi{10.1038/s41586-021-03878-5}

\bibitem[{{Lorimer} {et~al.}(2007){Lorimer}, {Bailes}, {McLaughlin},
  {Narkevic}, \& {Crawford}}]{Lorimer07}
{Lorimer}, D.~R., {Bailes}, M., {McLaughlin}, M.~A., {Narkevic}, D.~J., \&
  {Crawford}, F. 2007, Science, 318, 777, \dodoi{10.1126/science.1147532}

\bibitem[{{Lorimer} \& {Kramer}(2004)}]{Lorimer04handbook}
{Lorimer}, D.~R., \& {Kramer}, M. 2004, {Handbook of Pulsar Astronomy}, Vol.~4

\bibitem[{{Lu} {et~al.}(2022){Lu}, {Beniamini}, \& {Kumar}}]{Lu22}
{Lu}, W., {Beniamini}, P., \& {Kumar}, P. 2022, \mnras, 510, 1867,
  \dodoi{10.1093/mnras/stab3500}

\bibitem[{{Manchester} {et~al.}(2005){Manchester}, {Hobbs}, {Teoh}, \&
  {Hobbs}}]{Manchester05}
{Manchester}, R.~N., {Hobbs}, G.~B., {Teoh}, A., \& {Hobbs}, M. 2005, \aj, 129,
  1993, \dodoi{10.1086/428488}

\bibitem[{{Nimmo} {et~al.}(2023){Nimmo}, {Hessels}, {Snelders}, {Karuppusamy},
  {Hewitt}, {Kirsten}, {Marcote}, {Bach}, {Bansod}, {Barr}, {Behrend},
  {Bezrukovs}, {Buttaccio}, {Feiler}, {Gawro{\'n}ski}, {Lindqvist}, {Orbidans},
  {Puchalska}, {Wang}, {Winchen}, {Wolak}, {Wu}, \& {Yuan}}]{Nimmo23}
{Nimmo}, K., {Hessels}, J.~W.~T., {Snelders}, M.~P., {et~al.} 2023, \mnras,
  520, 2281, \dodoi{10.1093/mnras/stad269}

\bibitem[{{Niu} {et~al.}(2021){Niu}, {Li}, {Luo}, {Wang}, {Yao}, {Zhang},
  {Zhu}, {Wang}, {Ye}, {Zhang}, {Niu}, {Tang}, {Duan}, {Krco}, {Dai}, {Feng},
  {Miao}, {Pan}, {Qian}, {Xue}, {Yuan}, {Yue}, {Zhang}, \& {Zhang}}]{Niu21}
{Niu}, C.-H., {Li}, D., {Luo}, R., {et~al.} 2021, \apjl, 909, L8,
  \dodoi{10.3847/2041-8213/abe7f0}

\bibitem[{{Nyland} {et~al.}(2017){Nyland}, {Young}, {Wrobel}, {Davis},
  {Bureau}, {Alatalo}, {Morganti}, {Duc}, {de Zeeuw}, {McDermid}, {Crocker}, \&
  {Oosterloo}}]{Nyland17}
{Nyland}, K., {Young}, L.~M., {Wrobel}, J.~M., {et~al.} 2017, \mnras, 464,
  1029, \dodoi{10.1093/mnras/stw2385}

\bibitem[{{Oosterloo} {et~al.}(2010){Oosterloo}, {Morganti}, {Crocker},
  {J{\"u}tte}, {Cappellari}, {de Zeeuw}, {Krajnovi{\'c}}, {McDermid},
  {Kuntschner}, {Sarzi}, \& {Weijmans}}]{Oosterloo10}
{Oosterloo}, T., {Morganti}, R., {Crocker}, A., {et~al.} 2010, \mnras, 409,
  500, \dodoi{10.1111/j.1365-2966.2010.17351.x}

\bibitem[{{Ould-Boukattine} {et~al.}(2024){Ould-Boukattine}, {Chawla},
  {Hessels}, {Cooper}, {Gawro{\'n}ski}, {Herrmann}, {Kirsten}, {Hewitt},
  {Konijn}, {Nimmo}, {Pleunis}, {Puchalska}, \& {Snelders}}]{Ould-Boukattine24}
{Ould-Boukattine}, O.~S., {Chawla}, P., {Hessels}, J.~W.~T., {et~al.} 2024,
  arXiv e-prints, arXiv:2410.17024, \dodoi{10.48550/arXiv.2410.17024}

\bibitem[{{Perley} \& {Butler}(2017)}]{Perley17}
{Perley}, R.~A., \& {Butler}, B.~J. 2017, \apjs, 230, 7,
  \dodoi{10.3847/1538-4365/aa6df9}

\bibitem[{{Snelders} {et~al.}(2023){Snelders}, {Nimmo}, {Hessels}, {Bensellam},
  {Zwaan}, {Chawla}, {Ould-Boukattine}, {Kirsten}, {Faber}, \&
  {Gajjar}}]{Snelders23}
{Snelders}, M.~P., {Nimmo}, K., {Hessels}, J.~W.~T., {et~al.} 2023, Nature
  Astronomy, 7, 1486, \dodoi{10.1038/s41550-023-02101-x}

\bibitem[{{Spitler} {et~al.}(2016){Spitler}, {Scholz}, {Hessels}, {Bogdanov},
  {Brazier}, {Camilo}, {Chatterjee}, {Cordes}, {Crawford}, {Deneva}, {Ferdman},
  {Freire}, {Kaspi}, {Lazarus}, {Lynch}, {Madsen}, {McLaughlin}, {Patel},
  {Ransom}, {Seymour}, {Stairs}, {Stappers}, {van Leeuwen}, \&
  {Zhu}}]{Spitler16}
{Spitler}, L.~G., {Scholz}, P., {Hessels}, J.~W.~T., {et~al.} 2016, \nat, 531,
  202, \dodoi{10.1038/nature17168}

\bibitem[{{Strader} {et~al.}(2011){Strader}, {Romanowsky}, {Brodie}, {Spitler},
  {Beasley}, {Arnold}, {Tamura}, {Sharples}, \& {Arimoto}}]{Strader11}
{Strader}, J., {Romanowsky}, A.~J., {Brodie}, J.~P., {et~al.} 2011, \apjs, 197,
  33, \dodoi{10.1088/0067-0049/197/2/33}

\bibitem[{{Suresh} {et~al.}(2021){Suresh}, {Chatterjee}, {Cordes}, \&
  {Crawford}}]{Suresh21}
{Suresh}, A., {Chatterjee}, S., {Cordes}, J.~M., \& {Crawford}, F. 2021, \apj,
  920, 16, \dodoi{10.3847/1538-4357/ac1672}

\bibitem[{Virtanen {et~al.}(2020)Virtanen, Gommers, Oliphant, {et~al.}}]{SciPy}
Virtanen, P., Gommers, R., Oliphant, T.~E., {et~al.} 2020, Nature Methods, 17,
  261, \dodoi{10.1038/s41592-019-0686-2}

\bibitem[{{Wang} \& {Zhang}(2019)}]{Wang19_FRBDis}
{Wang}, F.~Y., \& {Zhang}, G.~Q. 2019, \apj, 882, 108,
  \dodoi{10.3847/1538-4357/ab35dc}

\bibitem[{{Wang} {et~al.}(2019){Wang}, {Lu}, {Zhang}, {Chen}, {Luo}, \&
  {Xu}}]{Wang19}
{Wang}, W., {Lu}, J., {Zhang}, S., {et~al.} 2019, Science China Physics,
  Mechanics, and Astronomy, 62, 979511, \dodoi{10.1007/s11433-018-9334-y}

\bibitem[{{White} \& {Becker}(1992)}]{White92_M49}
{White}, R.~L., \& {Becker}, R.~H. 1992, \apjs, 79, 331, \dodoi{10.1086/191656}

\bibitem[{{Xu} {et~al.}(2022){Xu}, {Niu}, {Chen}, {Lee}, {Zhu}, {Dong},
  {Zhang}, {Jiang}, {Wang}, {Xu}, {Zhang}, {Fu}, {Filippenko}, {Peng}, {Zhou},
  {Zhang}, {Wang}, {Feng}, {Li}, {Brink}, {Li}, {Lu}, {Yang}, {Caballero},
  {Cai}, {Chen}, {Dai}, {Djorgovski}, {Esamdin}, {Gan}, {Guhathakurta}, {Han},
  {Hao}, {Huang}, {Jiang}, {Li}, {Li}, {Li}, {Li}, {Li}, {Liu}, {Luo}, {Men},
  {Niu}, {Peng}, {Qian}, {Song}, {Stern}, {Stockton}, {Sun}, {Wang}, {Wang},
  {Wang}, {Wang}, {Wu}, {Xiao}, {Xiong}, {Xu}, {Xu}, {Yang}, {Yang}, {Yao},
  {Yi}, {Yue}, {Yu}, {Yu}, {Yuan}, {Zhang}, {Zhang}, {Zhang}, {Zhao}, {Zheng},
  {Zhu}, \& {Zou}}]{Xu22}
{Xu}, H., {Niu}, J.~R., {Chen}, P., {et~al.} 2022, \nat, 609, 685,
  \dodoi{10.1038/s41586-022-05071-8}

\bibitem[{{Zhang}(2018)}]{Zhang18}
{Zhang}, B. 2018, \apjl, 867, L21, \dodoi{10.3847/2041-8213/aae8e3}

\bibitem[{{Zhang}(2023)}]{Zhang23_FRBModel}
---. 2023, Reviews of Modern Physics, 95, 035005,
  \dodoi{10.1103/RevModPhys.95.035005}

\bibitem[{{Zhang} {et~al.}(2024{\natexlab{a}}){Zhang}, {Yang}, {Geng}, {Yang},
  \& {Wu}}]{Zhang24_highB}
{Zhang}, S.~B., {Yang}, X., {Geng}, J.~J., {Yang}, Y.~P., \& {Wu}, X.~F.
  2024{\natexlab{a}}, \apjl, 976, L26, \dodoi{10.3847/2041-8213/ad92fb}

\bibitem[{{Zhang} {et~al.}(2024{\natexlab{b}}){Zhang}, {Wang}, {Yang}, {Li},
  {Geng}, {Tang}, {Chang}, {Luo}, {Wang}, {Wu}, {Dai}, \&
  {Zhang}}]{Zhang24_M81}
{Zhang}, S.~B., {Wang}, J.~S., {Yang}, X., {et~al.} 2024{\natexlab{b}}, Nature
  Communications, 15, 7454, \dodoi{10.1038/s41467-024-51711-0}

\bibitem[{{Zhang} {et~al.}(2024{\natexlab{c}}){Zhang}, {Geng}, {Wang}, {Yang},
  {Kaczmarek}, {Tang}, {Johnston}, {Hobbs}, {Manchester}, {Wu}, {Jiang},
  {Huang}, {Zou}, {Dai}, {Zhang}, {Li}, {Yang}, {Dai}, {Chang}, {Pan}, {Lu},
  {Wei}, {Li}, {Wu}, {Qian}, {Wang}, {Wang}, {Feng}, \&
  {Staveley-Smith}}]{Zhang24_1913}
{Zhang}, S.~B., {Geng}, J.~J., {Wang}, J.~S., {et~al.} 2024{\natexlab{c}},
  \apj, 972, 59, \dodoi{10.3847/1538-4357/ad6602}

\bibitem[{{Zhang} {et~al.}(2022){Zhang}, {Wang}, {Feng}, {Zhang}, {Li}, {Tsai},
  {Niu}, {Luo}, {Yao}, {Zhu}, {Han}, {Lee}, {Zhou}, {Niu}, {Jiang}, {Wang},
  {Zhang}, {Xu}, {Wang}, \& {Xu}}]{Zhang22RAA}
{Zhang}, Y.-K., {Wang}, P., {Feng}, Y., {et~al.} 2022, RAA, 22, 124002,
  \dodoi{10.1088/1674-4527/ac98f7}

\bibitem[{{Zhu} {et~al.}(2023){Zhu}, {Xu}, {Zhou}, {Lin}, {Wang}, {Wang},
  {Zhang}, {Niu}, {Chen}, {Li}, {Meng}, {Lee}, {Zhang}, {Feng}, {Ge},
  {G{\"o}{\u{g}}{\"u}{\c{s}}}, {Guan}, {Han}, {Jiang}, {Jiang}, {Kouveliotou},
  {Li}, {Miao}, {Miao}, {Men}, {Niu}, {Wang}, {Wang}, {Xu}, {Xu}, {Xue},
  {Yang}, {Yu}, {Yuan}, {Yue}, {Zhang}, \& {Zhang}}]{Zhu23}
{Zhu}, W., {Xu}, H., {Zhou}, D., {et~al.} 2023, Science Advances, 9, eadf6198,
  \dodoi{10.1126/sciadv.adf6198}

\end{thebibliography}
\bibliographystyle{aasjournal}
%\clearpage



\end{document}